\documentclass[prb,reprint,amsmath,amssymb,showpacs,superscriptaddress,floatfix]{revtex4-2}
\usepackage[breaklinks=true,colorlinks,citecolor=blue,linkcolor=blue,urlcolor=blue]{hyperref}
\usepackage{epsfig,mathrsfs,color,latexsym,subfigure,marginnote,gensymb,}
\usepackage{graphicx}
\usepackage{xcolor}
\usepackage{chemformula}
\usepackage{multirow}
\usepackage{tabularx}
\usepackage[utf8]{inputenc}
\usepackage{makecell}
\renewcommand{\BibitemShut}[1]{}

\begin{document}
\title{Accelerating the prediction of stacking fault energy by combining ab initio calculations and machine learning}
\author{Albert Linda}
\affiliation{Department of Materials Science and Engineering, Indian Institute of Technology Kanpur, Kanpur 208016, India}
\author{Md. Faiz Akhtar}
\affiliation{Department of Materials Science and Engineering, Indian Institute of Technology Kanpur, Kanpur 208016, India}
\author{Shaswat Pathak}
\affiliation{Department of Mechanical Engineering, SRM College of Engineering And Technology, Kattankulathur-Chennai, 603203, India}
\author{Somnath Bhowmick}
\email[]{bsomnath@iitk.ac.in}
\affiliation{Department of Materials Science and Engineering, Indian Institute of Technology Kanpur, Kanpur 208016, India}

\date{\today}
\begin{abstract}
Stacking fault energies (SFEs) are key parameters to understand the deformation mechanisms in metals and alloys, and prior knowledge of SFEs from \textit{ab initio} calculations is crucial for alloy designing. Machine learning (ML) algorithms used in the present work show a $\sim$80 times acceleration of generalized stacking fault energy (GSFE) predictions, which are otherwise computationally very expensive to get directly from density functional theory (DFT) calculations, particularly for alloys. The origin of the features used for training the ML algorithms lies in the physics-based Friedel model, and the present work uncovers the connection between the physics of d-electrons and the deformation behavior of transition metals and alloys.  Predictions based on the ML model agree with the experimental data. Our model can be helpful in accelerated alloy designing by providing a fast method of screening materials in terms of stacking fault energies. 
\end{abstract}

\maketitle

\begin{figure*}
\includegraphics[width=\linewidth]{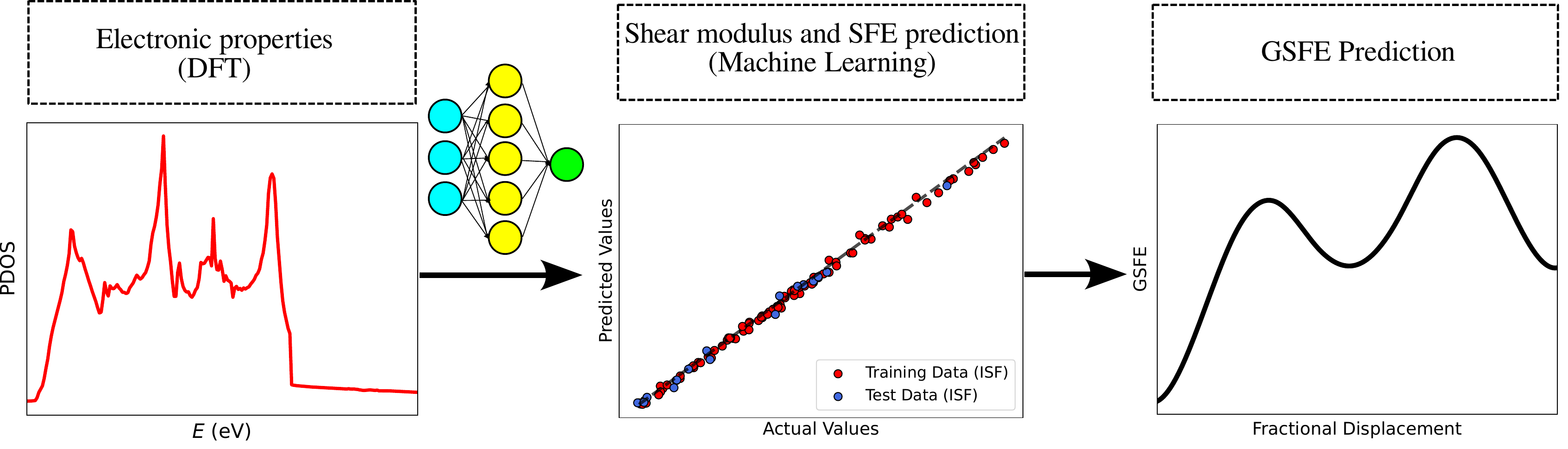}
\caption{Workflow for physics based accelerated generalized stacking fault energy (GSFE) calculation: \textit{ab initio} electronic density of states (DOS) calculation, followed by machine learning based prediction of stacking fault energy and GSFE curve.}
\label{fig:workflow}
\end{figure*}

\section{Introduction}
Stacking fault (SF) in face-centered cubic (FCC) materials is a planar defect that arises during plastic deformation through dissociating a perfect dislocation into two Shockley partial dislocations. Stacking fault energy (SFE) is a crucial parameter that determines the deformation mechanisms of FCC materials. Materials with low-to-medium SFE generally deform via transformation-induced plasticity (TRIP) or twinning-induced plasticity (TWIP), while those with high SFE deform via dislocation slip. SFE depends on several parameters like temperature~\cite{MOLNAR2019490, Zhang2018} and stress~\cite{ANDRIC2019262,linda2022effect} and it can be tuned via alloying~\cite{SHAO2017601, KUMAR2023101707, ZHANG2019807, ZHAO2017334}. Since SFE dictates dislocation dissociation, it is one of the determining factors for the dislocation pile-up at the twin boundaries (TBs), resulting in fatigue cracking~\cite{LI2023101011}. Deformation processing (like ball milling, rolling, and torsion) or lattice mismatch-induced interface strain can form high-density SFs in low-to-medium SFE metals, leading to strain hardening while maintaining good ductility~\cite{SU2021140696}. SFE plays a major role in the mechanical properties of bulk nanostructured materials processed via severe plastic deformation~\cite{AN20191}. The creep life of Ni-based superalloys improves due to SFE reduction by alloying with Co~\cite{TIAN2014316}. Due to its importance in the mechanical behavior of metals, several experimental and computational methods have been developed for SFE estimation, as discussed below. 

Experimentally, SFEs are estimated by transmission electron microscopy (TEM) or by X-ray diffraction (XRD) and neutron diffraction (ND). Using TEM, the intrinsic SFE is estimated by measuring the stacking fault width, which is defined as the separation distance of isolated pairs of leading and trailing partial dislocations. This method assumes a balance between the excess energy stored in the stacking fault and the elastic strain energy responsible for the mutual repulsion of leading and trailing partials~\cite{DELAVIGNETTE196217}. The determination of SFE through XRD and ND involves analyzing the shift and broadening of the Bragg peak, considering the relationship between stacking fault probability, dislocation density, and intrinsic SFE~\cite{Werner2021}. An \textit{in situ} XRD method to measure the critical stress in the early stage of plastic deformation provides another way of estimating SFE experimentally~\cite{Rafajanb5111,BYUN20033063}. 

The experimental methods mentioned above have one limitation - SFE at any unstable point (lying between perfect and faulted crystal) cannot be estimated. Such curves with SFE values at multiple points between perfect and faulted crystals are known as the generalized stacking fault energy (GSFE) profile or $\gamma$ surface. Computational methods like DFT or classical molecular dynamics (MD) are used to calculate the $\gamma$ surface~\cite{su2019density, hunter2014core, hu2013basal, linda2022effect, wu2010ab, JARLOV2022164137}. The $\gamma$ surface represents the potential energy landscape between adjacent planes in a slip system. Simulated $\gamma$ surface acts as an input for calculating the Peierls stresses via the Peierls-Nabarro model (P-N model) for studying dislocations~\cite{joos1994peierls, hartford1998peierls, shang2012temperature, shang2012effects, el2013effect, albert23, KAMIMURA2018355, albert23, XU2020102689, MA2022117447}, plastic deformation in high entropy alloys~\cite{Schonecker2021, ZHU2023119230} and phase transitions~\cite{Yang2023, WEN2022143011}. Due to its \textit{ab initio} nature, $\gamma$ surface predicted by DFT is believed to be very accurate, and the SFE values are in reasonable agreement with experimental findings. However, DFT calculation predicts negative SFEs for some materials like metastable alloys, which are experimentally reported to have small but positive SFE~\cite{Pei2021, WERNER2023101708, YOU2023103770, WEI2020992, Shih2021, Walter2020, CHANDAN2021113891}. Several attempts have been made to understand the reasons behind the discrepancy, further establishing the reliability of DFT for SFE prediction~\cite{Werner2021, SUN2021109396}.

Accuracy and reliability of DFT for SFE prediction lies in its ability to accurately incorporate the effect of electronic contributions~\cite{HU201696, SHI2018853, Stange2015, Wang2014, LI2022107556, Qi2007, Natarajan2020}. For example, I. R. Harris et al. showed the connection between the electronic structure (empty d-states) and SFE~\cite{harris1966influence}. Datta et al. found that the electronic density of states (DOS) plots for the faulted structures are considerably smoother compared to the pristine materials~\cite{DATTA20082531}. A study on the influence of solute substitutions in Ni on its GSFE found a correlation between density of state (DOS) and intrinsic stacking fault (ISF) energy~\cite{kumar2018influence}. The energy barriers for both deformation slip and twinning formation decrease with the increased electron concentrations in ZnS, ZnTe, and CdTe~\cite{Shen2020}. A recent study also revealed a direct correlation of SFE with the width of the d-band of FCC transition metals~\cite{linda2022effect}. As suggested by the previous studies, a deep connection exists between the electronic band structure and SFE, which we would like to explore in detail in the present work.

In contemporary times, machine learning (ML) algorithms have emerged as practical tools capable of achieving robust predictive outcomes for a given input dataset. Recent reports highlight the application of ML in several domains of materials science and engineering, like potential development~\cite{PhysRevLett.114.096405}, microstructure modeling~\cite{PhysRevMaterials.7.083802}, and structure-property correlation~\cite{PhysRevB.89.205118, ZHANG2020803}. In alloy development, ML has been employed for predicting phase stability, glass forming ability, and properties as a function of alloy composition~\cite{LIU2020182, KRISHNA2021113804, REVI2021110671}. Stacking fault energy, the subject matter of this paper has also been predicted using ML models using local composition, atomic size, electronic structure, physical, thermomechanical, and elastic properties as descriptors~\cite{Chong_2021, Mahato_2024, LIU2023171547, HU2021116800, WANG2021110544,KHAN2022117472}. However, it is noteworthy that the values of these fundamental properties for alloys are often estimated using the rule of mixture, introducing potential discrepancies in the results. A few studies have attempted to predict SFE using charge density obtained from DFT calculations~\cite{Aorra2022, ARORA2022101620}.

The novelty of the present work lies in its use of the physics-based Friedel model for deriving the features for machine learning. The physics-based model helps us to uncover the connection between the SFE and electronic band structure of FCC transition metals and alloys. A schematic diagram is illustrated in Figure~\ref{fig:workflow}. First, we calculate the electronic density of states (DOS), a routine job for DFT packages. Using the electronic DOS data, we calculate some parameters like the width of the d-band (W$_d$), energy at the band center ($\varepsilon_d$), electrons in the d-orbital (z$_d$), and electrons in the s-orbital (z$_s$). Using various machine learning models [Gaussian process regression (GPR), support vector regression (SVR), deep neural network (DNN), and random forest], we are able to predict the stacking fault energy and shear modulus of transition metals and alloys using the parameters obtained from DOS. Values predicted by the ML models agree with the experimental data. We are also able to predict the GSFE curve with reasonable accuracy, and our combined \textit{ab initio}-ML approach can accelerate the GSFE calculation 80x faster compared to solely \textit{ab initio} based approach in the case of alloys. Our work paves the way for fast and accurate computational prediction of transition metal alloys with desired SFE values, providing a valuable understanding of the deformation mechanism and mechanical behavior. 

\begin{figure}[htbp]
\includegraphics[width=\columnwidth]{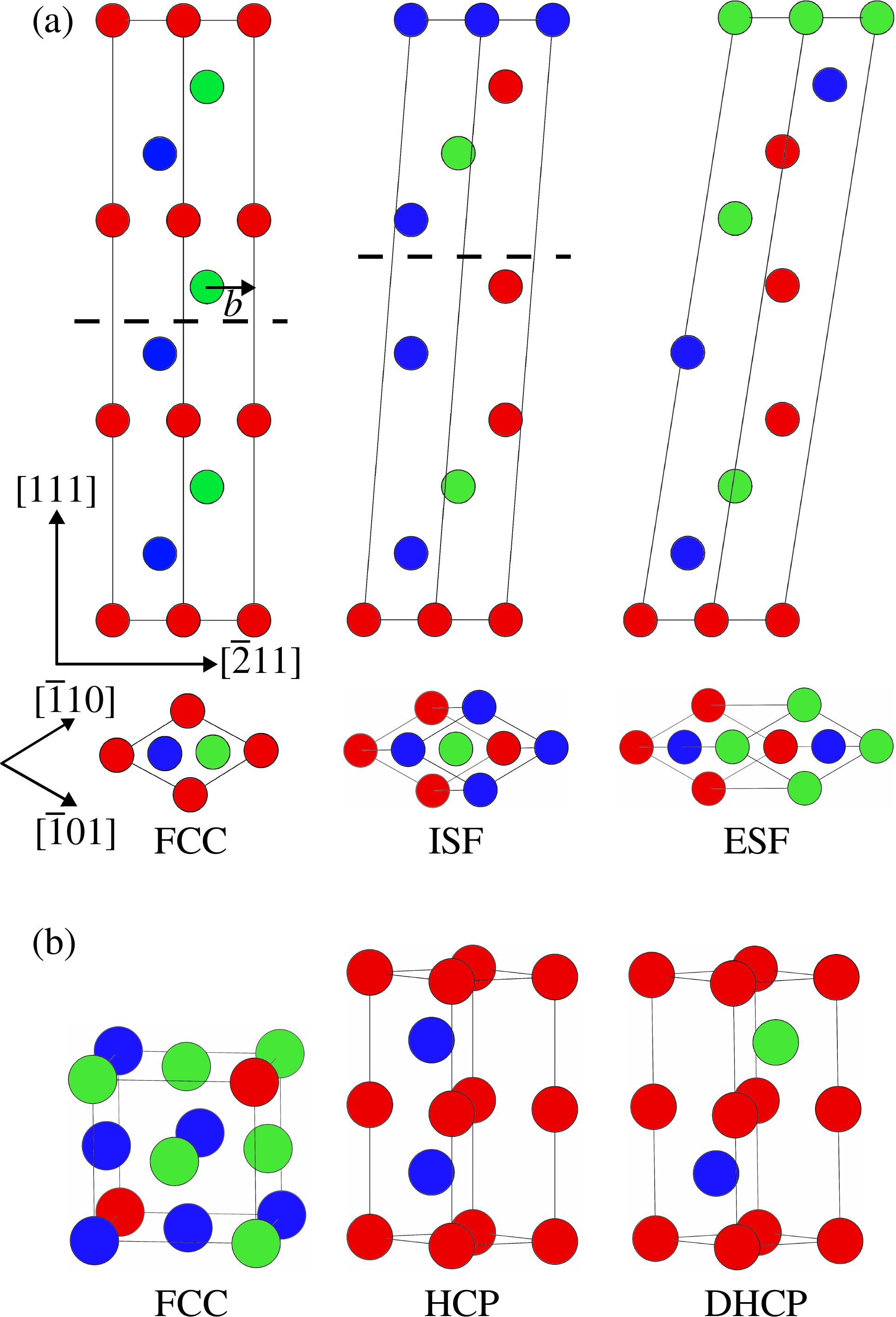}
\caption{(a) Supercell method: side view (first row) and top view (second row) of the supercell of the face centered cubic or FCC (left), intrinsic stacking fault or ISF (center) and extrinsic stacking fault or ESF (right). Starting from the FCC structure, all the atoms located above the dotted line and the out-of-plane cell vector are displaced by $\vec{b}$ (2$\vec{b}$) to go from the FCC to ISF (ESF) structure. (b) ANNNI model: FCC, hexagonal closed packed or HCP and double hexagonal closed packed or DHCP cells used for the stacking fault energy calculations. In both (a) and (b) the A, B and C stacking sequence of atoms along the closed packed direction are represented in red, blue and green colors respectively.}
\label{fig:structure_supercell_annni}
\end{figure}

\section{Methodology}
Density functional theory (DFT) calculations, as implemented in the Vienna Ab-initio Simulation Package (VASP) \cite{PhysRevB.59.1758}, are performed using a plane wave basis set (with a 400 eV kinetic energy cut-off) and projector augmented wave (PAW) potentials \cite{PhysRevB.54.11169}. The generalized gradient approximation (GGA), applying Perdew, Burke, and Ernzerhof (PBE) as exchange-correlation functional~\cite{PhysRevLett.77.3865}, is used. The unit cell parameters and atomic coordinates are fully relaxed until the energy converges to within $10^{-6}$ eV and the atomic force dips below 0.01 eV/\AA. Further details about the supercell size and k-point mesh used for Brillouin zone sampling are given in the respective sections.

\begin{figure}[htbp]
\includegraphics[width=\columnwidth]{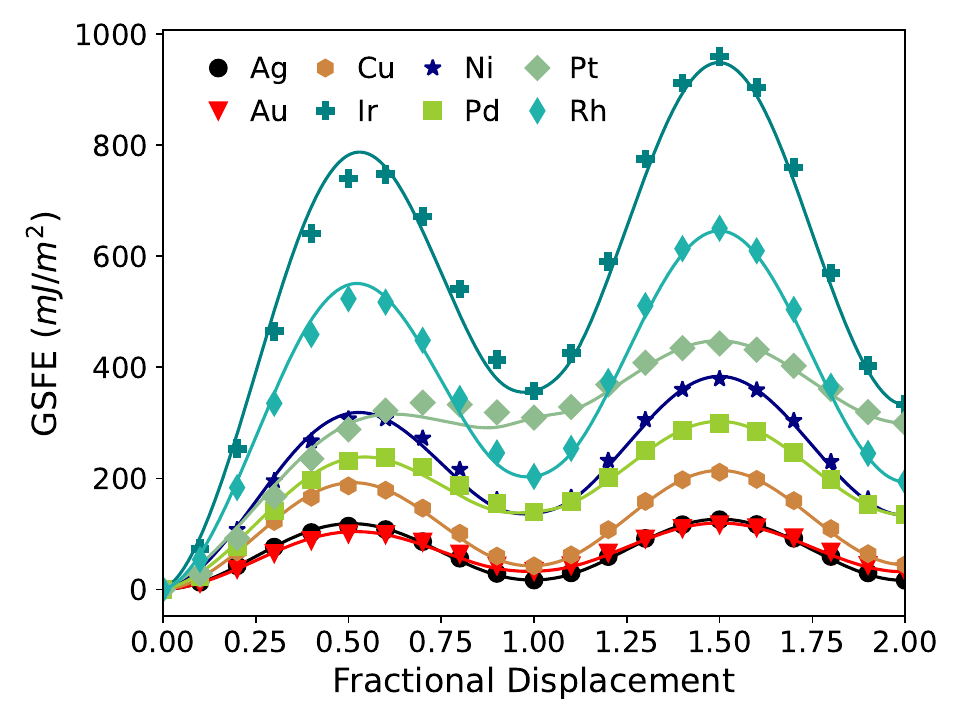}
\caption{Generalized stacking fault energy (GSFE) curves for FCC metals, illustrated along the $[\overline{2}11]$ direction. Energy values are plotted from 0 to $2\Vec{b}$, where Bergers vector $\Vec{b}=\frac{1}{6}[\overline{2}11]$. Symbols depict DFT values, while the curves are fitted using Equation~\ref{eq:gsfe_curve}.}
\label{fig:gsfe_metals}
\end{figure}

\section{Results and discussion}
\subsection{Stacking fault energy calculations}
\begin{table*}[htbp]
\renewcommand{\arraystretch}{1.5} 
\caption{Comparison of SFEs obtained from DFT using two approaches (supercell and ANNNI model), predicted SFEs (using deep neural network or DNN) and experimental data~\cite{sfe_1965}$^a$, \cite{sfe_1970}$^b$, \cite{sfe_1983}$^c$. A similar comparison for shear modulus values obtained through DFT, predicted using DNN and experimental values \cite{shear_mod_metals}$^d$. Both DFT and ML-predicted values are in good agreement with the experimental values.}
\begin{center}
\begin{tabular}{|c|ccccc|ccc|cc|}
\hline
\textbf{Metals} & \multicolumn{5}{c|}{\textbf{DFT}}                                                                                                                                                                                       & \multicolumn{3}{c|}{\textbf{Predicted}}                                                                                                                     & \multicolumn{2}{c|}{\textbf{Exp.}}                                                    \\ \hline
                           & \multicolumn{2}{c|}{\textbf{Supercell}}                                                     & \multicolumn{2}{c|}{\textbf{ANNNI}}                                                         & \multirow{2}{*}{\textbf{G}} & \multicolumn{1}{c|}{\multirow{2}{*}{\textbf{$\gamma_{ISF}$}}} & \multicolumn{1}{c|}{\multirow{2}{*}{\textbf{$\gamma_{ESF}$}}} & \multirow{2}{*}{\textbf{G}} & \multicolumn{1}{c|}{\multirow{2}{*}{\textbf{$\gamma_{ISF}$}}} & \multirow{2}{*}{\textbf{G$^d$}} \\ \cline{1-5}
                           & \multicolumn{1}{c|}{\textbf{$\gamma_{ISF}$}} & \multicolumn{1}{c|}{\textbf{$\gamma_{ESF}$}} & \multicolumn{1}{c|}{\textbf{$\gamma_{ISF}$}} & \multicolumn{1}{c|}{\textbf{$\gamma_{ESF}$}} &                             & \multicolumn{1}{c|}{}                                         & \multicolumn{1}{c|}{}                                         &                             & \multicolumn{1}{c|}{}                                         &                             \\ \hline
    Ag                     & \multicolumn{1}{c|}{16.9}                    & \multicolumn{1}{c|}{16.3}                    & \multicolumn{1}{c|}{17.5}                    & \multicolumn{1}{c|}{18.4}                    & 22.0                        & \multicolumn{1}{c|}{18.1}                                     & \multicolumn{1}{c|}{23.2}                                     & 22.8                        & \multicolumn{1}{c|}{25.0$^a$}    &                         27.0    \\ \hline
    Au                     & \multicolumn{1}{c|}{32.6}                    & \multicolumn{1}{c|}{31.7}                    & \multicolumn{1}{c|}{23.6}                    & \multicolumn{1}{c|}{23.3}                    & 15.4                        & \multicolumn{1}{c|}{32.8}                                     & \multicolumn{1}{c|}{37.5}                                     & 19.2                        & \multicolumn{1}{c|}{45.0$^a$}    &                         27.7    \\ \hline
    Cu                     & \multicolumn{1}{c|}{42.4}                    & \multicolumn{1}{c|}{44.6}                    & \multicolumn{1}{c|}{48.7}                    & \multicolumn{1}{c|}{53.3}                    & 49.8                        & \multicolumn{1}{c|}{43.8}                                     & \multicolumn{1}{c|}{56.3}                                     & 39.6                        & \multicolumn{1}{c|}{55.0$^b$}    &                         48.3    \\ \hline
    Ir                     & \multicolumn{1}{c|}{357.2}                   & \multicolumn{1}{c|}{333.1}                   & \multicolumn{1}{c|}{348.3}                   & \multicolumn{1}{c|}{334.3}                   & 214.4                       & \multicolumn{1}{c|}{359.6}                                    & \multicolumn{1}{c|}{400.3}                                    & 216.1                       & \multicolumn{1}{c|}{480.0$^c$}   &                         210.0    \\ \hline
    Ni                     & \multicolumn{1}{c|}{136.6}                   & \multicolumn{1}{c|}{133.9}                   & \multicolumn{1}{c|}{140.8}                   & \multicolumn{1}{c|}{135.0}                   & 95.1                        & \multicolumn{1}{c|}{138.5}                                    & \multicolumn{1}{c|}{162.7}                                    & 95.1                        & \multicolumn{1}{c|}{125.0$^c$}   &                         75.0    \\ \hline
    Pd                     & \multicolumn{1}{c|}{139.5}                   & \multicolumn{1}{c|}{134.3}                   & \multicolumn{1}{c|}{146.6}                   & \multicolumn{1}{c|}{139.5}                   & 44.4                        & \multicolumn{1}{c|}{137.1}                                    & \multicolumn{1}{c|}{148.3}                                    & 45.1                        & \multicolumn{1}{c|}{130.0$^a$}   &                         43.6    \\ \hline
    Pt                     & \multicolumn{1}{c|}{309.1}                   & \multicolumn{1}{c|}{299.5}                   & \multicolumn{1}{c|}{277.0}                   & \multicolumn{1}{c|}{282.6}                   & 48.6                        & \multicolumn{1}{c|}{299.8}                                    & \multicolumn{1}{c|}{315.6}                                    & 50.6                        & \multicolumn{1}{c|}{322.0$^c$}   &                         61.0    \\ \hline
    Rh                     & \multicolumn{1}{c|}{203.4}                   & \multicolumn{1}{c|}{194.3}                   & \multicolumn{1}{c|}{190.2}                   & \multicolumn{1}{c|}{188.2}                   & 146.8                       & \multicolumn{1}{c|}{207.0}                                    & \multicolumn{1}{c|}{240.4}                                    & 150.5                       & \multicolumn{1}{c|}{330.0$^a$}   &                         150.0    \\ \hline
    Pd-Pt                  & \multicolumn{1}{c|}{190.8}                   & \multicolumn{1}{c|}{180.9}                   & \multicolumn{1}{c|}{176.0}                   & \multicolumn{1}{c|}{172.0}                   & 45.8                        & \multicolumn{1}{c|}{172.2}                                    & \multicolumn{1}{c|}{186.5}                                    & 47.0                        & \multicolumn{1}{c|}{-}                                        &        -                     \\ \hline
    Ir-Pt                  & \multicolumn{1}{c|}{359.5}                   & \multicolumn{1}{c|}{342.9}                   & \multicolumn{1}{c|}{328.5}                   & \multicolumn{1}{c|}{326.2}                   & 163.5                       & \multicolumn{1}{c|}{326.6}                                    & \multicolumn{1}{c|}{371.5}                                    & 150.5                       & \multicolumn{1}{c|}{-}                                        &        -                     \\ \hline
    Pd-Au                  & \multicolumn{1}{c|}{131.4}                   & \multicolumn{1}{c|}{128.2}                   & \multicolumn{1}{c|}{118.0}                   & \multicolumn{1}{c|}{112.0}                   & 37.4                        & \multicolumn{1}{c|}{116.0}                                    & \multicolumn{1}{c|}{123.9}                                    & 37.5                        & \multicolumn{1}{c|}{-}                                        &        -                     \\ \hline
\end{tabular}
\end{center}
\label{tbl:sfe}
\end{table*}

\subsubsection{SFE using periodic supercell}
We consider an ideal FCC structure composed of 9 layers stacked in an ...ABCABCABC... pattern [Figure~\ref{fig:structure_supercell_annni}(a)]. Two of the cell vectors, $\frac{1}{2}[\overline{1}10]$ and $\frac{1}{2}[\overline{1}01]$, lie on the (111) plane, while the third one is perpendicular to the (111) plane and aligned along the [111] direction. An intrinsic stacking fault (ISF) has a stacking sequence of ...ABCABABCABC..., as shown in Figure~\ref{fig:structure_supercell_annni}(a). 

An ISF is created by fixing the bottom five layers and displacing each of the top four layers by the Burgers vector $\vec{b}=\frac{1}{6}[\overline{2}11]$. Simultaneously, we shift the out-of-the-plane cell vector (oriented initially along the [111] direction) by the same vector $\vec{b}$ to preserve the unit cell's periodicity. This approach enables us to compute the stacking fault energy using periodic cells, eliminating the need for introducing surface layers~\cite{Kibey2007}. We define the intrinsic stacking fault energy $\gamma_{ISF}$ as the energy difference between the faulted and ideal structures per unit area:
\begin{equation}
\gamma_{ISF}=\frac{E_{ISF}-E_{FCC}}{A}.
\label{eq_gammasf}
\end{equation}
To get the energy values for metals from DFT calculations, we use  $21\times 21\times 2$ k-point mesh. 

An extrinsic stacking fault (ESF) has a stacking sequence of ...ABCABACABC..., as shown in  Figure~\ref{fig:structure_supercell_annni}(a). Starting with the ISF structure, we now fix the bottom six layers and displace the top three layers by $\vec{b}=\frac{1}{6}[\overline{2}11]$. The out-of-the-plane cell vector is also shifted by $\vec{b}$, yielding the ESF stacking sequence [Figure~\ref{fig:structure_supercell_annni}(a)]. Notably, in the case of ESF, the top 3 layers and the out-of-the-plane cell vector are displaced by $2\vec{b}$ relative to the ideal FCC configuration. To determine the $\gamma_{ESF}$ value, we employ an expression similar to Equation~\ref{eq_gammasf}.

Apart from the $\gamma_{ISF}$ and $\gamma_{ESF}$, we compute stacking fault energies at various displacements, ranging from 0 to $2b$, with a step size of $0.1b$ to delineate the entire GSFE curve, as shown in Figure \ref{fig:gsfe_metals}. Two significant peaks along the GSFE curve are noteworthy—one situated at approximately the middle of ideal FCC and ISF (referred to as the unstable stacking fault or USF), and the other located at approximately the middle of ISF and ESF (referred to as the unstable twinning fault or UTF). These peaks represent the energy barriers for forming ISF and ESF, respectively.

We illustrate the GSFE curves of all the transition and noble metals having FCC ground state in Figure \ref{fig:gsfe_metals}. The symbols in the figure represent the values calculated from DFT. The curves are drawn using the following expressions:  
\begin{equation}
    \small
    \gamma = 
\begin{cases}
    c G \sin^2(\pi x) + \gamma_{ISF}\cdot x, & 0 \leq x \leq 1 \\
    c G \sin^2(\pi x) + \gamma_{ISF} \cdot (2 - x) + \gamma_{ESF} \cdot (x - 1),  & 1 \leq x \leq 2
\end{cases}
\label{eq:gsfe_curve}
\end{equation}
where $\gamma_{ISF}$, $\gamma_{ESF}$ and shear modulus $G$ are calculated from DFT, $c$ is a constant, and $x$ is the fractional displacement in terms of Burgers vector $\Vec{b}$. The values of $c$ are  5.01 for Ag, 5.67 for Au, 3.43 for Cu, 3.74 for Pd, 2.96 for Pt, 2.61 for Ni,  3.04 for Rh, and 2.81 for Ir. We calculate the shear modulus via strain-energy approach~\cite{Page2001} by using VASPKIT tool~\cite{WANG2021108033}, details of which are given in Section I, Supplemental Material (SM)~\cite{supp}. In conclusion, one can generate the entire GSFE curve with reasonable accuracy by calculating three numbers: $\gamma_{ISF}$, $\gamma_{ESF}$, and $G$ from DFT. Such an approach is computationally cheaper than calculating the entire GSFE curve from DFT, particularly when dealing with alloys.

\subsubsection{SFE using ANNNI model}
Axial-next-nearest-neighbor ising (ANNNI) model is an alternate route for finding SFEs. Although the model is computationally less expensive, one can get only the ISF and ESF values instead of the entire GSFE curve. ANNNI model uses specific combinations of energies corresponding to different short-period stacking sequences of close-packed (111) planes. For example, the second-order approximation to obtain the ISF and ESF energies is given by the following combinations:
\begin{eqnarray}
    E_{ISF}=\frac{E_{HCP}+2E_{DHCP}-3E_{FCC}}{A},\\\nonumber
    E_{ESF}=\frac{4(E_{DHCP}-E_{FCC})}{A}.
    \label{eq:annni}
\end{eqnarray}
In the above equation, $A=\frac{\sqrt{3}}{4}a^2$, where $a$ is the lattice parameter of a conventional FCC unit cell. Energies of the face-centered cubic (ABCABC stacking), hexagonal close-packed (ABAB stacking), and double hexagonal close-packed (ABACABAC stacking) structures are denoted by $E_{\text{FCC}}$, $E_{\text{HCP}}$, and $E_{\text{DHCP}}$, respectively. FCC, HCP, and DHCP unit cells used for the SFE calculation using the ANNNI model are illustrated in Figure~\ref{fig:structure_supercell_annni}(b). To get the energy values for metals from DFT calculations, we use  $12\times 12\times 12$, $21\times 21\times 5$, and $21\times 21\times 5$ k-point mesh for FCC, HCP, and DHCP, respectively. SFE values calculated from the supercell method and ANNNI model are compared in Table~\ref{tbl:sfe}. Besides Au, values obtained from both models are in remarkable agreement.

\begin{figure*}[htbp]
\centering
\includegraphics[width=\linewidth]{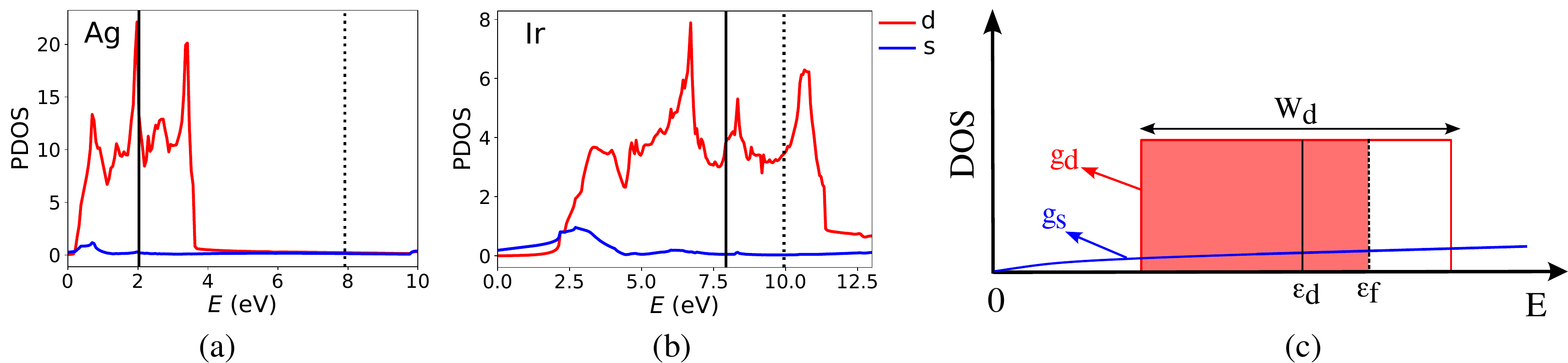}
\caption{Orbital projected density of states for (a) Ag (completely full narrow d-band, low SFE) and (b) Ir (partially full wide d-band, high SFE). The black solid line denotes the d-band center ($\varepsilon_d$), while the dotted line represents the Fermi energy ($E_{F}$). (c) Schematic s-band and d-band electronic density of states, according to the Freidel model. }
\label{fig3}
\end{figure*}

\subsection{Friedel model}
Understanding electronic structure is the primary building block for a comprehensive study of the material's properties. Electrons serve as the quantum glue that keeps the nuclei of a solid together and influences the mechanical, electrical, optical, and magnetic properties of materials. It is well known that d-electrons play a significant role in transition metals' electronic and magnetic properties. Figures~\ref{fig3}(a)-(b) illustrate the DOS of s-electrons and d-electrons of Ag and Ir, respectively. Unlike the DOS of s-states, the DOS of d-states is sharply peaked, which indicates that d-states are relatively localized compared to the s-states. Although the DOS curves are quite intricate, Friedel proposed a significant simplification. The DOS of s-states, denoted by $g_s(\varepsilon)$, is approximated to be free electron like, obeying $g_s(\varepsilon) \propto \sqrt{\varepsilon}$ [Figure~\ref{fig3}(c)]. The DOS of d-states, denoted by $g_d(\varepsilon)$, is approximated to be a step function [Figure~\ref{fig3}(c)], expressed as,
\begin{equation}
\begin{aligned}g_d\left( \varepsilon \right) =\dfrac{10}{W_{d}},  \hspace{1.0cm}\varepsilon _{d}-\dfrac{W_{d}}{2} <\varepsilon <\varepsilon _{d}+\dfrac{W_{d}}{2}\\ =0.         \hspace{3.40cm}     \rm{otherwise}.\end{aligned}
\label{eqdos}
\end{equation}
The center of the d-band and its width are denoted by $\varepsilon_{d}$ and $W_d$, which are related to the projected density of states (PDOS) of the d-band. The first moment of DOS with respect to the Fermi energy ($\varepsilon _{F}$) is,
\begin{equation}
\mu=\int (\varepsilon -\varepsilon_F)g^{DFT}_d(\varepsilon) d\varepsilon,   
\end{equation}
where $g^{DFT}_d(\varepsilon)$ is the PDOS of d-band, obtained from the \textit{ab initio} calculations. The number $\varepsilon_d=(\varepsilon_F-|\mu|)$ corresponds to the center of the d-band. Further, we calculate the second moment of the DOS with respect to $\varepsilon_d$,
\begin{equation}
\sigma^2=\int (\varepsilon -\varepsilon_d)^2g^{DFT}_d(\varepsilon)d\varepsilon.   
\label{eqwd}
\end{equation}
We define the width of the d-band as $W_d$ = $2\sigma$. As shown in Figure S1 and S2 in SM, the periodic trend of calculated $\varepsilon_d$ and $W_d$ agree with the solid-state table~\cite{harrison2012electronic}.

It is evident that, unlike s-states, d-states can not be treated using free electron theory, and a tight binding-like description would be more appropriate. In a tight binding description, bandwidth is an important parameter that depends on the overlap of atomic orbitals. For example, core states have zero width because of no overlap. Valence d-states have a finite width, leading to some energy gain, depending on $W_d$. Using the DOS expression in Equation~\ref{eqdos}, one can illustrate that the energy gain is,
\begin{equation}
    E_{d}=5W_{d}\left[ -\dfrac{z_{d}}{10}+\left(\dfrac{z_{d}}{10}\right) ^{2}\right],
    \label{eq:ed}
\end{equation}
where $z_d$ is the number of electrons in the d-band. We compute $z_d$ from the \textit{ab initio} calculations by integrating the d-band PDOS up to the Fermi energy. We obtain $W_d$ from \textit{ab initio} calculations using Equation~\ref{eqwd}. We define $E_d$ as the cohesive energy due to the overlap of adjacent d-bands. The term within the square bracket in Equation~\ref{eq:ed} has a minimum at $z_d=5$ (middle of the transition metal series), and it is zero at $z_d=10$ (noble metal). Our \textit{ab initio} calculations confirm that $z_d$ increases as we move from left to right of a row in the periodic table [Figure S3 in SM]. However, $z_d$ is slightly less than 10 in noble metals, as some electrons are transferred to the free electron-like band. Interestingly, we also find a periodic trend in $W_d$ along a particular row; values increase from the left to the center and decrease from the center to the noble metal. In other words, $W_d$ has a maximum near the middle of the transition metal series [Figure S2 in SM]. According to the Friedel model, the binding energy [Equation~\ref{eq:ed}] of transition metals is maximum near the middle of a row [Figure S4 in SM]. This trend is in reasonably good agreement with experimental values. For example, the melting point is higher near the middle of the transition metal series [Figure S4 in SM]. Such a correlation makes the Friedel model credible despite its simplicity. 

We calculate the Wigner-Seitz radius $r_0$ by equating volume per atom (obtained from \textit{ab initio}) to $4\pi r_0^3/3$. Values of $r_0$ obtained from \textit{ab initio} agree well with the ones reported in the solid state table [Figure S5, SM]. Since d-band overlap decreases with increasing distance between the atoms, we assume bandwidth $W_d\propto r_0^{-\alpha}$. As a result, volume dependence of $E_d$, denoted by $\kappa_d$, can be expressed as,
\begin{equation}
    \kappa_d=\frac{\partial E_d}{\partial r_0}=\frac{5\alpha W_{d}}{r_0}\left[\dfrac{z_{d}}{10}-\left( \dfrac{z_{d}}{10}\right) ^{2}\right].
\end{equation}
Similar to $E_d$, $\kappa_d$ also peaks near the middle of the transition metal series [Figure S6, SM].

In summary, the Friedel model defines binding among d-electrons in terms of specific material parameters, which can be computed from the electronic density of states obtained from \textit{ab initio} calculations. In the following section, we use these parameters to fit a machine learning model, which can predict SFE values of transition metals and binary alloys.

\subsection{Machine learning}
\begin{figure}[htbp]
\centering
\includegraphics[width=\columnwidth]{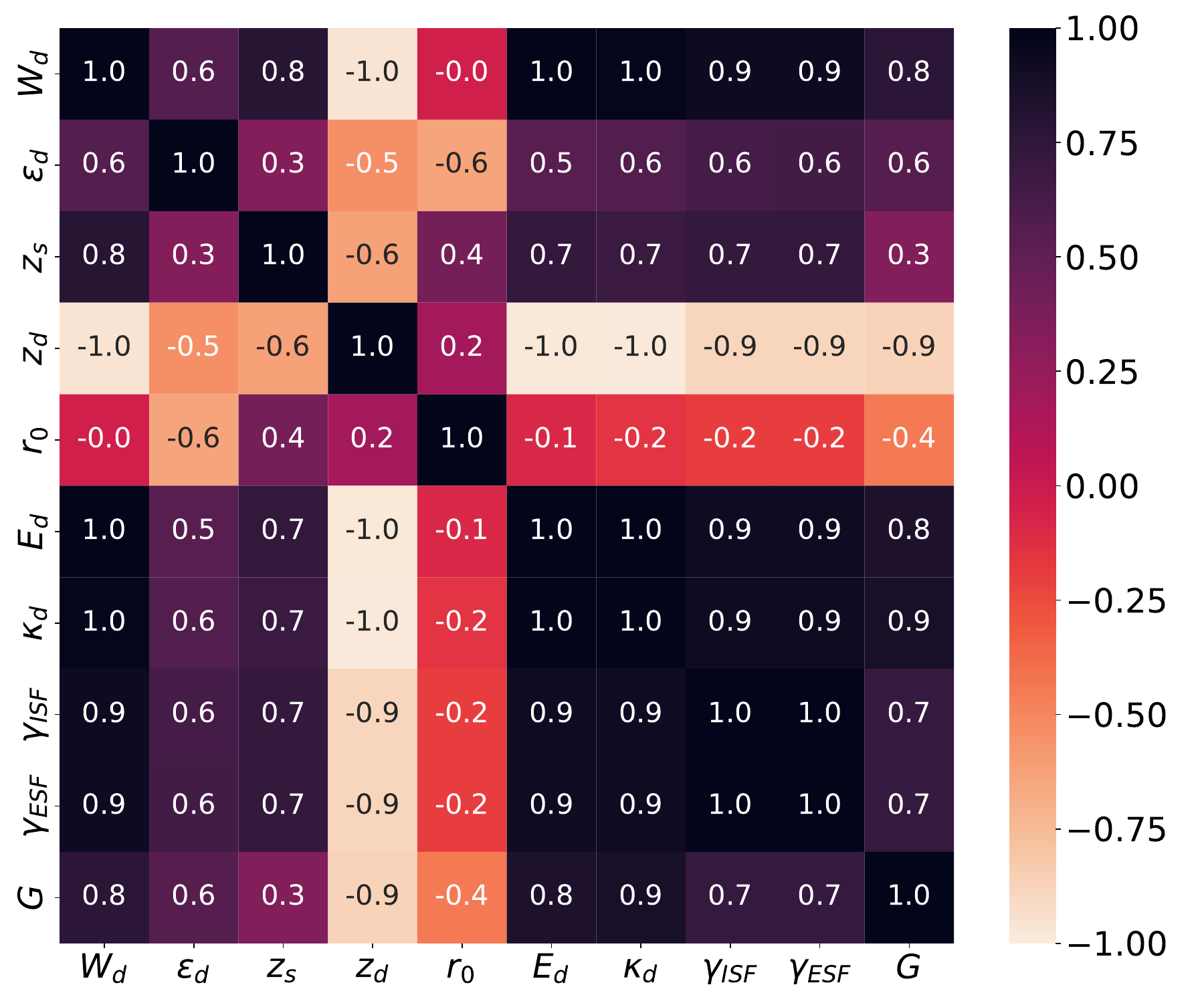}
\caption{Heatmap of Pearson correlation coefficient matrix: target variables are $\gamma_{ISF},\gamma_{ESF}, G$ and feature variables are $W_d, \varepsilon_d, z_s, z_d, r_0, E_d, \kappa_d$. Values close to black (white) indicate strongly positive (negative) correlations.}
\label{fig:correlation}
\end{figure}
\subsubsection{Data generation using ANNNI model}
We use the ANNNI model to generate an extensive database of $\gamma_{ISF}$ and $\gamma_{ESF}$ values for Au-Pd, Pd-Ag, Ag-Au, Rh-Pd, Ir-Pd, Pd-Pt, Cu-Pt, Ir-Pt, Ni-Ag, Ni-Au, Ni-Pd, Ni-Pt, Ni-Rh and Ni-Cu binary alloys. These alloys are selected because of the solid solubility of the two elements throughout the composition range, spanning from 12.5\% to 87.5\%, with intervals of 12.5\%, encompassing seven compositions for each alloy. Using the ATAT package~\cite{VANDEWALLE201313}, we generate special quasirandom structures (SQS) to describe the random arrangement of constituent atoms in a binary alloy. We generate three types of supercells, each containing 32 atoms: a conventional $2\times 2\times 2$ FCC supercell, a $2\times 2\times 4$ HCP supercell, and a $2\times 2\times 2$ DHCP supercell. We use a k-point mesh of $12\times 12\times 12$ for FCC, $21\times 21\times 5$ for HCP and DHCP supercell. The complete dataset for training and testing the ML model contains $\gamma_{ISF}$, $\gamma_{ESF}$ and $G$ values for 8 metals and all the binary alloys mentioned above. We also calculate the d-band PDOS and related parameters [Figure~\ref{fig3}(c)] using the FCC supercell of the metals and alloys.

\subsubsection{Feature and model selection} 
We aim to train a model to predict $\gamma_{ISF}, \gamma_{ESF}$ and $G$ of a material from its DOS, such that one can generate the GSFE curves using Equation~\ref{eq:gsfe_curve}. For the purpose of prediction, we use $\varepsilon_d, W_d, z_d, r_0$, and $z_s$ (number of s-electrons) as feature variables. Except $r_0$, the rest of the features have moderate to high values of correlation coefficients [Figure~\ref{fig:correlation}]. Notably, $z_d$ (number of d-electrons) has a very high negative correlation with SFEs, which implies lower SFE for a material with higher $z_d$. This observation agrees with the experimental facts that noble metals (Au, Ag, Cu) have low SFEs, as they have the highest d-electrons. Bandwidth $W_d$ has a very high positive correlation with SFEs, which is again consistent with the fact that noble metals have narrow bands compared to others [Figure~\ref{fig3} and Figure S7, S8 in SM], resulting in low SFEs.

Although some features have high correlation coefficients, a multivariable linear regression fails to predict the target variables accurately. Thus, we use other regression methods like deep neural network (DNN), support vector regression (SVR), Gaussian process regression (GPR), and random forest. We split the data set for training and testing (80:20). The latter is used to test the trained model and compute the test error.  The mean absolute error between the actual and predicted values gives the loss. We select the model that exhibits the highest coefficient of determination for total average $R^2$ for the test set and the highest total $R^2$ for the training set as the optimal one for each approach. The following discussion covers DNN and random forest, while SVR and GPR are given in Section II and Figure S9, SM.

\begin{figure*}
\centerline{\includegraphics[width=0.86\textwidth]{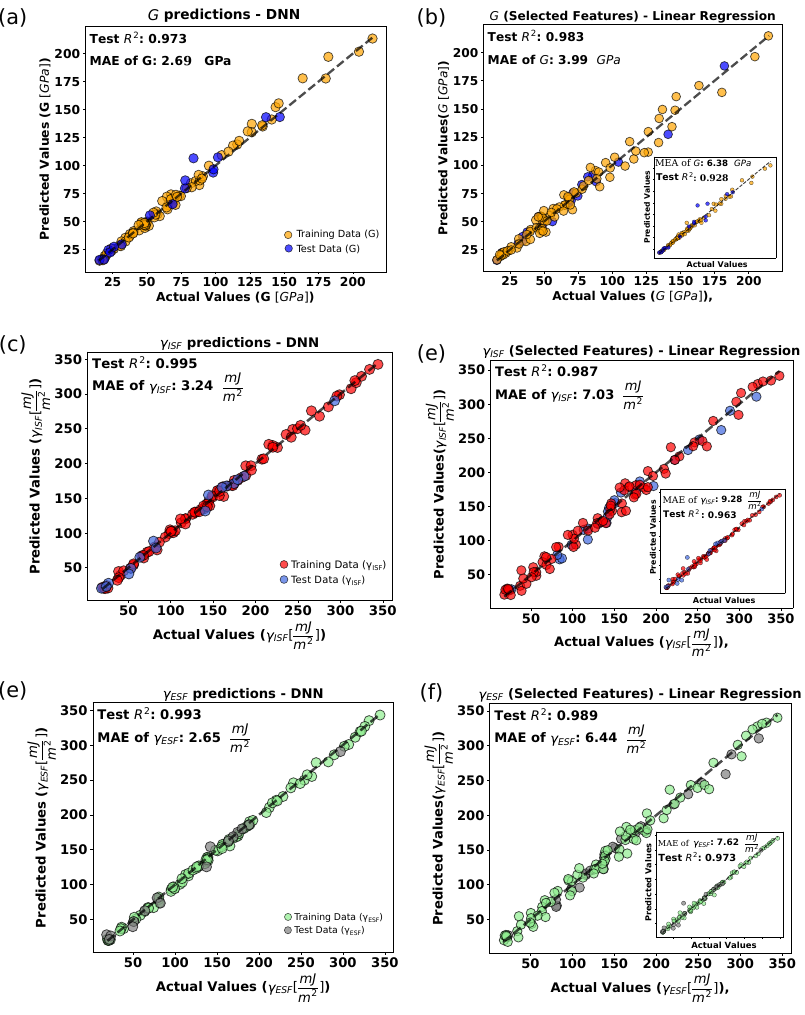}}
\caption{Panels (a), (c) and (e) illustrate the comparison between actual and predicted values from DNNs for $G$, $\gamma_{ISF}$ and  $\gamma_{ESF}$, respectively. Panels (b), (d), and (f) depict the same for selected features using Equations \ref{eqpg}, \ref{eqpisf} and \ref{eqpesf}, while the insets illustrate the outcomes of random forest regression with all the features.}
\label{figpredicted}
\end{figure*}

\subsubsection{Deep neural network} 
DNNs can capture highly non-linear relationships and complex patterns because of their highly flexible and expressive interconnected architecture~\cite{SCHMIDHUBER201585}. We evaluate the performance of different activation functions, like rectified linear unit (ReLU), leaky ReLU, and parametric ReLU (PReLU). PReLU demonstrates superior overall performance, achieving the highest accuracy among the tested activation functions with a test $R^2$ of 0.995 for $\gamma_{ISF}$ prediction, compared to leaky ReLU (0.993) and traditional ReLU (0.988). The neural network with PReLU activation showcases enhanced resistance to sample bias because of its adaptive nature to effectively modulate activation for negative inputs and minimize outliers' impact while promoting superior generalization for a more reliable and stable predictive model than leaky ReLU and ReLU counterparts. Figure~\ref{figpredicted} (a), (c), (e) shows the predicted vs. actual values for the best models of DNN, which we train with 5-7 dense layers with a learning rate of $10^{-3}$ with around 200-500 epochs for iterations. We evaluate the model's performance based on the test error and the change in loss with the iterations. Convergence with the number of iterations is shown in Figure S10, SM. The test $R^2$ values of $G$, $\gamma_{ISF}$ and$\gamma_{ESF}$ are 0.973, 0.995 and 0.993, respectively. The mean absolute errors (MAEs) of $G$, $\gamma_{ISF}$ and$\gamma_{ESF}$ are 2.69 GPa, 3.24 mJ/m$^2$ and 2.65 mJ/m$^2$, respectively.

\subsubsection{Random forest}
\begin{figure}[htbp]
\centering
\includegraphics[width=\columnwidth]{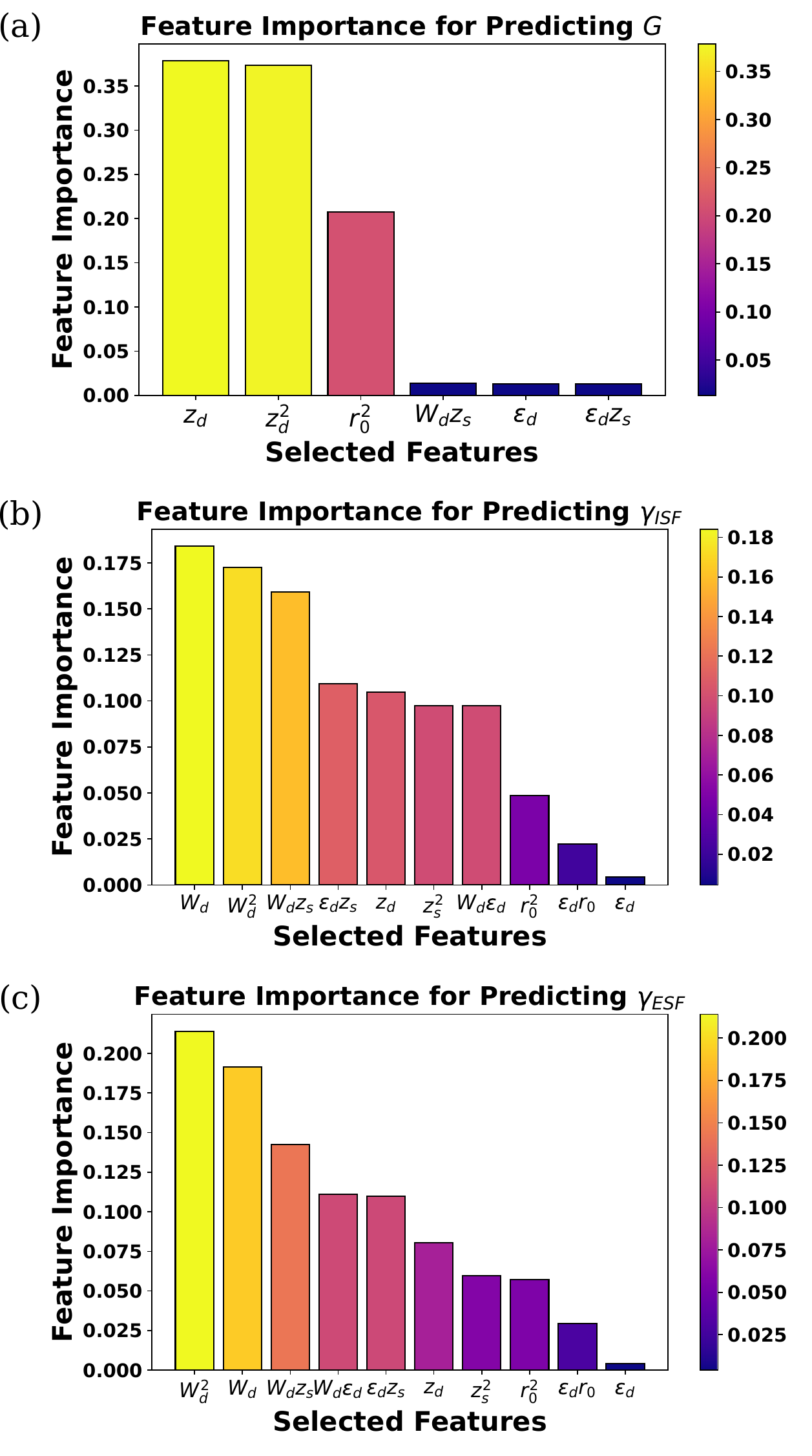}
\caption{Feature importance plot for $G$, $\gamma_{ISF}$, and $\gamma_{ESF}$.}
\label{fig:feature_importance}
\end{figure}
While the DNN exhibited impressive accuracy in predicting, it is not possible to understand how $\gamma_{ISF}, \gamma_{ESF}$, and $G$ depend on the feature variables. Our next objective is to predict the expression for $\gamma_{ISF}, \gamma_{ESF}$ and $G$ in terms of the feature variables. For this purpose, one must perform high-order polynomial regression, such as quadratic regression. This method expands the sample space from the initial five parameters ($\varepsilon_d, W_d, z_d, r_0$, and $z_s$) to twenty parameters by incorporating quadratic combinations. However, employing this approach may introduce redundant parameters, potentially leading to overfitting. 

A strategy to mitigate overfitting is to utilize a random forest regressor, incorporating the quadratic terms. The advantage of employing random forest lies in its ability to perform regression and simultaneously provide insights into the minimum number of terms essential for optimal prediction without the issue of overfitting by performing a search using its randomly ensembled decision trees. This process is called feature importance analysis. Details of feature importance analysis using random forest are given in Section III, SM. After feature importance analysis, we select only six terms for shear modulus prediction and ten terms for SFE prediction [Figure \ref{fig:feature_importance}]. Finally, we do a multivariable linear regression with the selected features to obtain the following expressions, which can be directly used for prediction. 
\begin{equation}
\begin{aligned}
G = 51.07  z_{d}^{2} - 63.21  r_{0}^{2} + 25.82  \varepsilon_{d}\\ - 876.15  z_{d} 
+ 33.80  W_{d}  z_{s} - 60.14  \varepsilon_{d}  z_{s} + 3877.91.
\end{aligned}
\label{eqpg}
\end{equation}
\begin{equation}
\begin{aligned}
\gamma_{\text{ISF}} = -22.66  W_{d}^{2} - 579.36  z_{s}^{2} - 522.06  r_{0}^{2} - 220.46  \varepsilon_{d}\\
+ 27.77  z_{d} + 64.57  W_{d} + 18.51  W_{d}  \varepsilon_{d} + 278.81  W_{d}  z_{s}\\
- 49.31  \varepsilon_{d}  z_{s} + 74.77  \varepsilon_{d}  r_{0} + 996.58.
\end{aligned}
\label{eqpisf}
\end{equation}                                            
\begin{equation}
\begin{aligned}
\gamma_{\text{ESF}} = -21.29  W_{d}^{2} - 730.97  z_{s}^{2} - 453.58  r_{0}^{2} - 178.69  \varepsilon_{d}\\
+ 31.77  z_{d} + 78.59  W_{d} + 16.30  W_{d}  \varepsilon_{d} + 249.45  W_{d}  z_{s}\\
+ 0.80  \varepsilon_{d}  z_{s} + 41.06  \varepsilon_{d}  r_{0} + 781.64.
\end{aligned}
\label{eqpesf}
\end{equation}
Note that, before applying the feature importance selection analysis, the mean absolute error obtained by including all the twenty terms are 6.38 GPa, 9.28 mJ/m$^2$, and 7.62 mJ/m$^2$ for G, $\gamma_{ISF}$ and $\gamma_{ESF}$ [insets of Figure~\ref{figpredicted} (b), (d), (f)], which reduces to 3.99 GPa, 7.03 mJ/m$^2$, and 6.44 mJ/m$^2$ [Figure~\ref{figpredicted} (b), (d), (f)], respectively. The test $R^2$ values of $G$, $\gamma_{ISF}$ and$\gamma_{ESF}$ also improve from  0.928, 0.963, and 0.973 (with all twenty features) to 0.983, 0.987 and 0.989 (with selected features), respectively. The improvement can be attributed to keeping only essential features, thus reducing the problem of overfitting.

Figure \ref{fig:feature_importance} illustrates all the selected features that are utilized in predicting the formula [Equation~\ref{eqpg},~\ref{eqpisf}, ~\ref{eqpesf}] in descending order in terms of their importance. Two features are dominant for shear modulus $G$: linear and quadratic terms of the number of d-electrons ($z_d$), followed by $r_0^2$. Stacking fault energies $\gamma_{ISF}$ and $\gamma_{ESF}$ depend on multiple features, the linear and quadratic term of d-band width ($W_d$) being the most important among them. The list also contains some cross terms like $W_d\varepsilon_d, W_dz_s, \varepsilon_dz_s$ with non-negligible weight, highlighting the highly non-linear nature of the problem, which requires a combined approach involving state-of-the-art \textit{ab initio} calculations and machine learning methods for complete understanding.

\subsection{GSFE curve prediction}
\begin{figure*}
\includegraphics[width=1.0\textwidth]{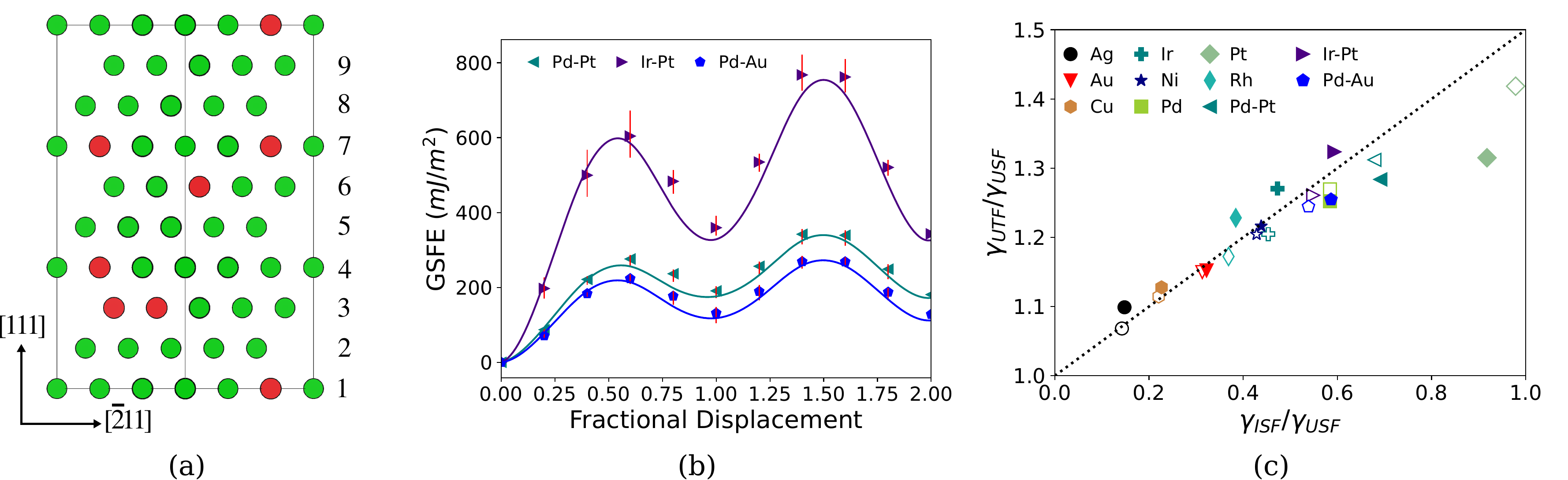}
\caption{(a) Supercell for GSFE calculation of alloys; numbers adjacent to the atomic planes represent the layer number. (b) GSFE curves for Pd$_{0.75}$Pt$_{0.25}$, Ir$_{0.75}$Pt$_{0.25}$, and Pd$_{0.75}$Au$_{0.25}$ alloys. The symbols are from DFT calculations, while the curves are generated from Equation~\ref{eq:gsfe_curve}, using $\gamma_{ISF}$, $\gamma_{ESF}$, and $G$ values predicted via ML. (c) $\gamma_{UTF}/\gamma_{USF}$ scales linearly with $\gamma_{ISF}/\gamma_{USF}$, as shown by the dotted line. Comparison of actual values from DFT calculations (solid symbols) and predicted values (open symbols).} 
\label{fig:gsfe_alloys_universal}
\end{figure*}
So far, we have focused on training ML models for predicting $\gamma_{ISF}, \gamma_{ESF}$ and $G$. Finally, we take up the most challenging task of predicting the entire GSFE. Conventionally, one should calculate the GSFE curves for several alloys using DFT and use them to train ML models. However, calculating the GSFE curves for alloys is computationally very expensive. Instead, we use the predicted $G$, $\gamma_{ISF}$ and $\gamma_{ESF}$ values from the previous section and construct the GSFE curves using Equation~\ref{eq:gsfe_curve}. For a binary alloy, we use the rule of mixture to get the value of $c$ (listed after Equation~\ref{eq:gsfe_curve}), which is the weighted average of the pure element's values. The following discussion shows that our method makes GSFE prediction 80X faster for alloys.

Figure~\ref{fig:gsfe_alloys_universal} compares the predicted GSFE curves with actual DFT values, illustrated for binary Pd-Pt, Ir-Pt, and Pd-Au alloys. We use the same technique as described earlier (Figure~\ref{fig:structure_supercell_annni}), but with a nine times larger supercell, having cell vectors $\frac{3}{2}[\overline{1}10], \frac{3}{2}[\overline{1}01], [111]$. Such a supercell contains eighty-one atoms, nine each in 9 different layers [Figure~\ref{fig:gsfe_alloys_universal}(a)]. We generate SQS to describe the random arrangement of constituent atoms in a binary alloy and use a k-point mesh of $9\times 9\times 3$. Because of the randomness, each layer has a different composition [Figure~\ref{fig:gsfe_alloys_universal}(a)], and the GSFE curve depends on the specific choice of layers during the deformation. For example, we start by fixing layer 1 and displacing layers 2 to 9, followed by fixing layers 1-2 and displacing layers 3 to 9, etc., as shown in Figure~\ref{fig:gsfe_alloys_universal}(a). Thus, we have to repeat the calculation eight times, and the average value yields one single DFT data point on a GSFE curve [Figure~\ref{fig:gsfe_alloys_universal}(b)]. The error bars show the lowest and highest among the eight DFT values calculated. Since there are 10 data points on a GSFE curve, we need to perform 80 calculations to get the entire GSFE curve from DFT directly. 

Considering the large number of atoms in the supercell, predicting GSFE directly from DFT is computationally expensive for alloys. As an alternative, the proposed ML approach requires only one DFT calculation to get the DOS and compute relevant parameters like $\varepsilon_d, W_d, z_d, z_s$, and $r_0$. Using these parameters, one can predict $\gamma_{ISF}, \gamma_{ESF}$, and $G$ using the ML model and finally predict the GSFE curve using Equation~\ref{eq:gsfe_curve}. Figure~\ref{fig:gsfe_alloys_universal}(b) illustrates that the ML-predicted GSFE curves are in good agreement with the actual DFT points (based on eighty DFT calculations). As shown in Figure~\ref{fig:gsfe_alloys_universal}(c), $\gamma_{UTF}/\gamma_{USF}$ scales linearly with $\gamma_{ISF}/\gamma_{USF}$. Predicted values agree reasonably well with the DFT results. 

\section{Conclusions}
In conclusion, we have proposed a combined \textit{ab initio} and ML-based model that can accelerate the computational prediction of GSFE curves for alloys by a factor of 80. The training dataset is generated using DFT calculations to find the SFE values of 106 metals and alloys using the ANNNI model. The features used for training the ML algorithms come from the physics-based Friedel model. The features are obtained from the electronic DOS, calculated using DFT. Other than accelerating the process of GSFE calculation, the present work also highlights a deep connection between the physics of d-electrons and the deformation behavior of transition metals and alloys. Our study reveals a highly non-linear dependence of shear modulus and stacking fault energies on the electronic features, which requires a combined approach involving state-of-the-art ab initio calculations and machine learning methods for complete understanding. The present model can accelerate alloy designing with targeted mechanical behavior by providing a fast method of screening materials in terms of stacking fault energies.

\section{Acknowledgements}
We acknowledge National Super Computing Mission (NSM) for providing computing resources of ``PARAM Sanganak'' at IIT Kanpur, which is implemented by CDAC and supported by the Ministry of Electronics and Information Technology (MeitY) and Department of Science and Technology (DST), Government of India. We also thank ICME National Hub, IIT Kanpur and CC, IIT Kanpur for providing HPC facility.

\bibliography{references_supplement}

\end{document}


\title{Accelerating the prediction of stacking fault energy by combining ab initio calculations and machine learning}

\author{Albert Linda}
\affiliation{Department of Materials Science and Engineering, Indian Institute of Technology Kanpur, Kanpur 208016, India}
\author{Md. Faiz Akhtar}
\affiliation{Department of Materials Science and Engineering, Indian Institute of Technology Kanpur, Kanpur 208016, India}
\author{Shaswat Pathak}
\affiliation{Department of Mechanical Engineering, SRM College of Engineering And Technology, Kattankulathur-Chennai, 603203, India}
\author{Somnath Bhowmick}
\email[]{bsomnath@iitk.ac.in}
\affiliation{Department of Materials Science and Engineering, Indian Institute of Technology Kanpur, Kanpur 208016, India}

\date{\today}

\maketitle

\section{Estimation of shear modulus}
Energy depends on strain via\cite{Page2001},
\begin{equation}
    \Delta E(V,\varepsilon_{i}) = E(V,\varepsilon_{i}) - E(V_0) = \frac{V_0}{2} \sum_{i,j=1}^{6} C_{ij} \varepsilon_{j} \varepsilon_{i},
    \label{eq:energy-strain}
\end{equation}
where $E(V,\varepsilon_{i})$ and $E(V_0)$ are the energies of distorted and ideal lattice respectively. During the application of strain the lattice vectors transform like
\begin{equation}
   \begin{bmatrix}
       a^{\prime} \\
       b^{\prime} \\
       c^{\prime}
   \end{bmatrix}
   = 
   \begin{bmatrix}
       a \\
       b \\
       c
   \end{bmatrix}(I + \epsilon).
   \label{eq:strained_lv}
\end{equation}
Here, $I$ represents the identity matrix, $a$, $b$, $c$ and $a^{\prime}$, $b^{\prime}$, $c^{\prime}$ are the lattice vectors of the undeformed and deformed structure respectively. The above equation requires strain matrix $\epsilon$, which is given by
\begin{equation}
   \epsilon =
   \begin{bmatrix}
       \varepsilon_{1} & \varepsilon_{6}/2 & \varepsilon_{5}/2 \\
       \varepsilon_{6}/2 & \varepsilon_{2} & \varepsilon_{4}/2 \\
       \varepsilon_{5}/2 & \varepsilon_{4}/2 & \varepsilon_{3} 
   \end{bmatrix}.
   \label{eq:strained_lv}
\end{equation}
For cubic material, three types of strain matrix is required each for estimating  $C_{44}$, $C_{11} + C_{12}$ and $C_{11} + 2C_{12}$. The matrices are obtained by utilizing the values given in the following table.
\begin{table}[h]
    \centering
    \begin{tabular}{|c|c|c|c|c|c|c|}  
        \hline  
        & $\varepsilon_{1}$ & $\varepsilon_{2}$ & $\varepsilon_{3}$ & $\varepsilon_{4}$ & $\varepsilon_{5}$ & $\varepsilon_{6}$ \\ \hline  
        $C_{44}$ & 0 & 0 & 0 & $\delta$ & $\delta$ & $\delta$ \\ \hline  
        $C_{11} + C_{12}$ & $\delta$ & $\delta$ & 0 & 0 & 0 & 0 \\ \hline   
        $C_{11} + 2C_{12}$ & $\delta$ & $\delta$ & $\delta$ & 0 & 0 & 0 \\ \hline  
    \end{tabular}
\end{table}

Also, we know that for cubic material the elastic stiffness matrix is given by 
\begin{equation}
   C_{ij} =
   \begin{bmatrix}
       C_{11} & C_{12} & C_{12} & 0 & 0 & 0 \\
       C_{12} & C_{11} & C_{12} & 0 & 0 & 0 \\
       C_{12} & C_{12} & C_{11} & 0 & 0 & 0 \\
       0 & 0 & 0 & C_{44} & 0 & 0 \\
       0 & 0 & 0 & 0 & C_{44} & 0 \\
       0 & 0 & 0 & 0 & 0 & C_{44} \\
   \end{bmatrix}.
\end{equation}
Using $C_{ij}$ and the strain matrix of all three cases, equation \ref{eq:energy-strain} change to 
\begin{equation}
    \Delta E = \frac{3}{2} C_{44} V \delta^2,
\end{equation}
\begin{equation}
    \Delta E = (C_{11} + C_{12})V \delta^2
\end{equation}
and 
\begin{equation}
    \Delta E = \frac{3}{2} (C_{11} + 2C_{12})V \delta^2.
\end{equation}
Using above three equation we can estimate $C_{11}$, $C_{12}$ and $C{44}$. These values can then be used to estimate the Voigt bound for shear modulus($G$). This is the upper bound\cite{voigt1910lehrbuch}, given by
\begin{equation}
    G_V = \frac{C_{11} - C_{12} + 3C_{44}}{5}.
\end{equation}
The Reuss bound is the lower bond\cite{Chandrasekar1989}, given by 
\begin{equation}
    G_{R} = \frac{5}{4(S_{11} - S_{12}) + 3S_{44}}.
\end{equation}
$S_{ij}$ values can be obtained using the following equations, 
\begin{eqnarray}
S_{11} = \frac{C_{11}+C_{12}}{(C_{11}-C_{12})(C_{11}+2C_{12})},\\
S_{12} = \frac{-C_{12}}{(C_{11}-C_{12})(C_{11}+2C_{12})},\\
S_{44} = \frac{1}{C_{44}}.
\end{eqnarray}
Based on the  Voigt and Reuss bounds, we can estimate the Voigt-Reuss-Hill average, which corresponds to the values for poly-crystalline material\cite{Hill1952}. This is given by
 \begin{equation}
     G_{VRH} = \frac{G_{V} + G_{R}}{2}.
 \end{equation}
We use VASPKIT\cite{WANG2021108033}, which implements the above algorithm. Strain values are taken to be -0.004, -0.002, 0.000, 0.002, and 0.004. We use $14 \times 14 \times 14$ k-point mesh for both metals and alloys to estimate energies for $G$ calculation.

\newpage

\begin{figure*}
    \centering
    \includegraphics[width=0.9\textwidth]{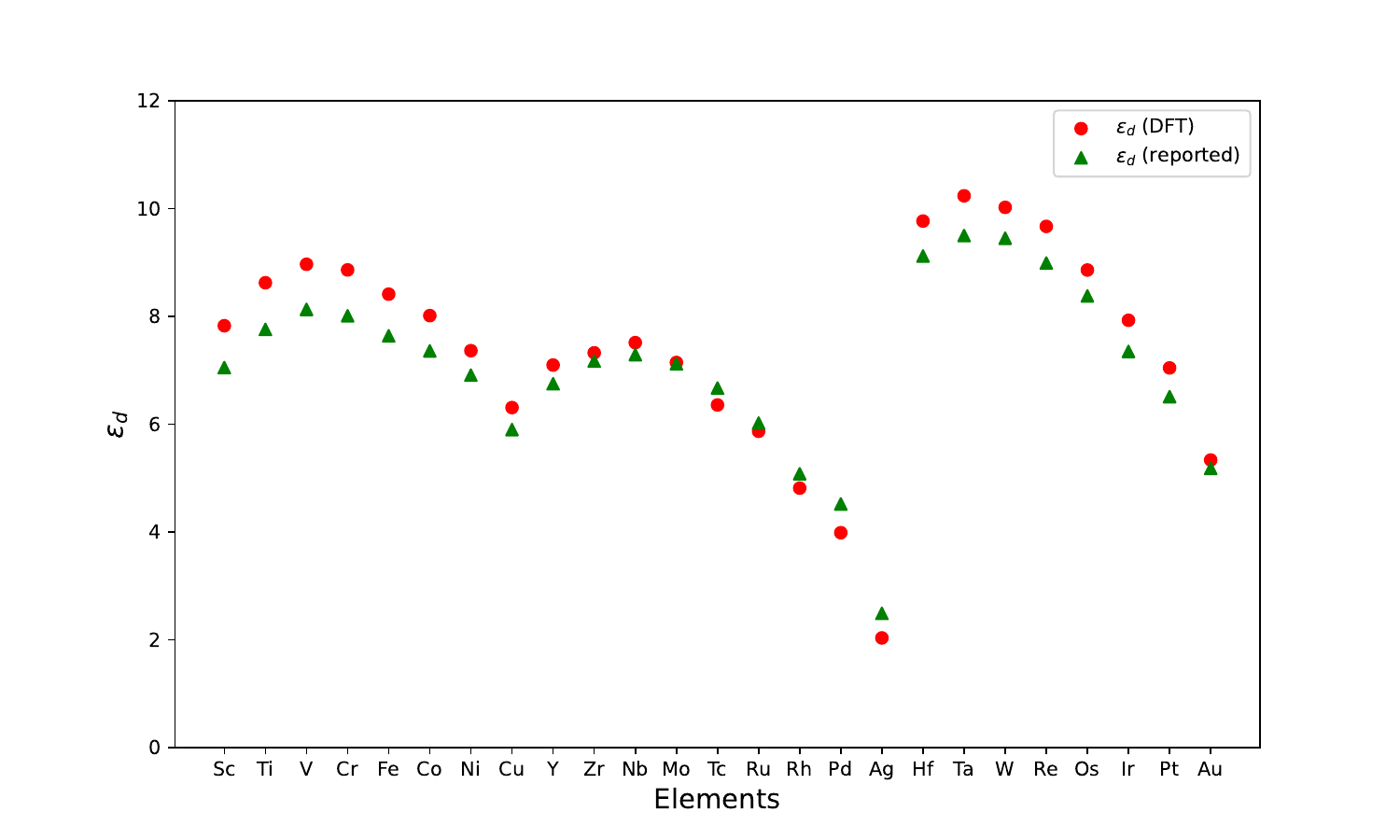}
    \caption{A comparison of d-band center $\varepsilon_{d}$ between reported values~\cite{harrison2012electronic} and DFT values for 3d, 4d and 5d elements. }
\end{figure*}

\begin{figure*}
    \centering
    \includegraphics[width=0.9\textwidth]{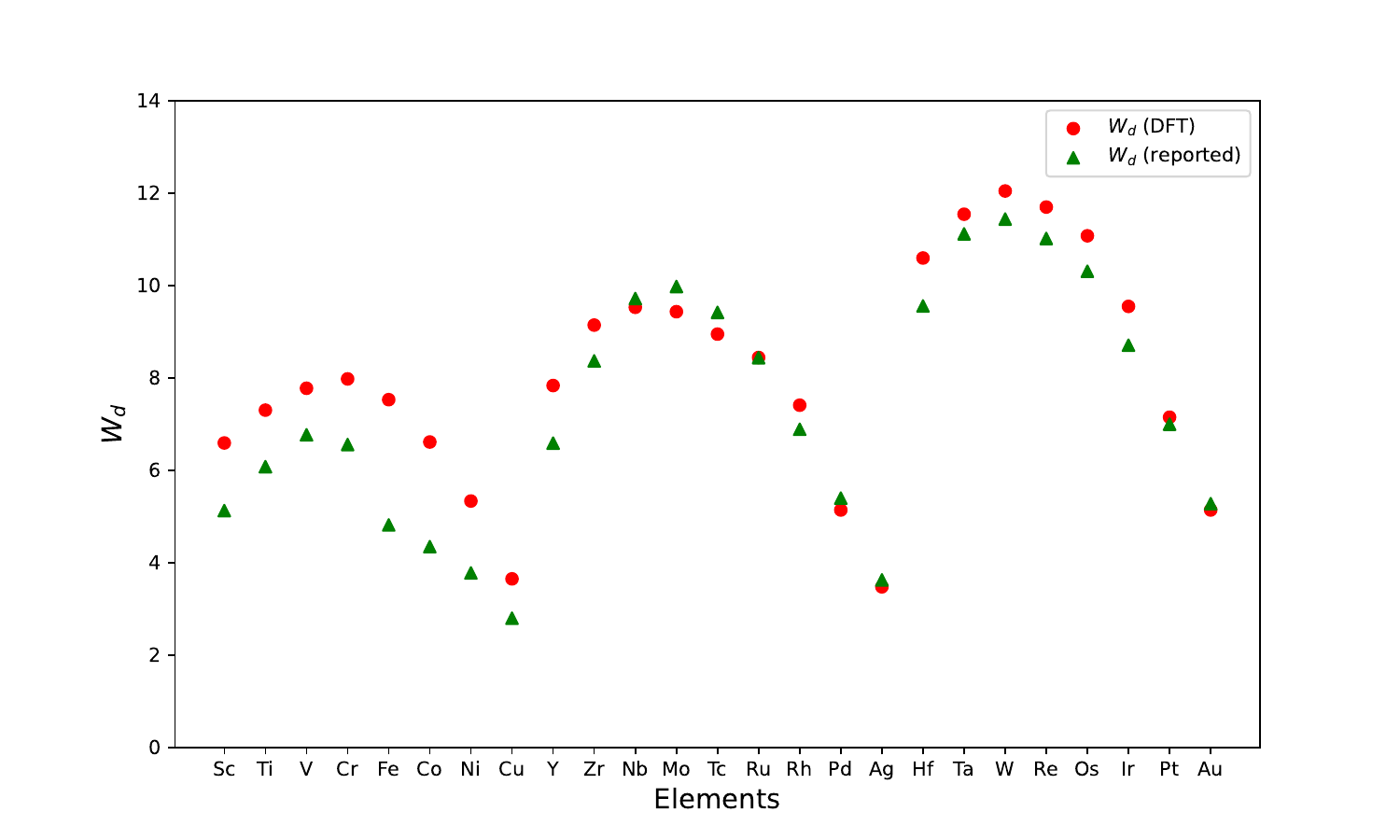}
    \caption{A comparison of d-band width $W_d$ between reported values~\cite{harrison2012electronic} and DFT values for 3d, 4d and 5d elements.}
\end{figure*}

\begin{figure*}
    \centering
    \includegraphics[width=0.9\textwidth]{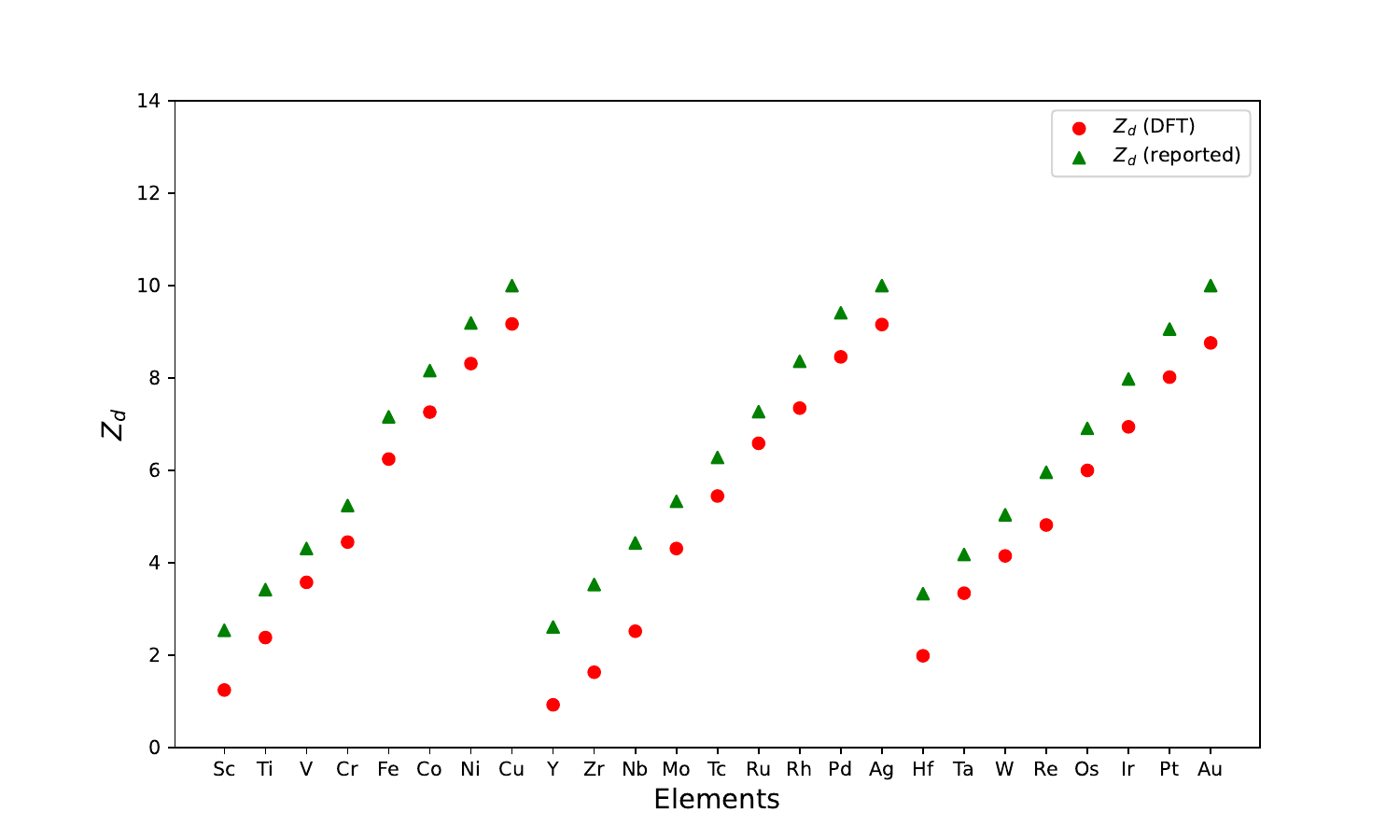}
    \caption{A comparison of valance d-electrons $z_{d}$ between reported values~\cite{harrison2012electronic} and DFT values for 3d, 4d and 5d elements.}
\end{figure*}

\begin{figure*}
    \centering
    \includegraphics[width=0.9\textwidth]{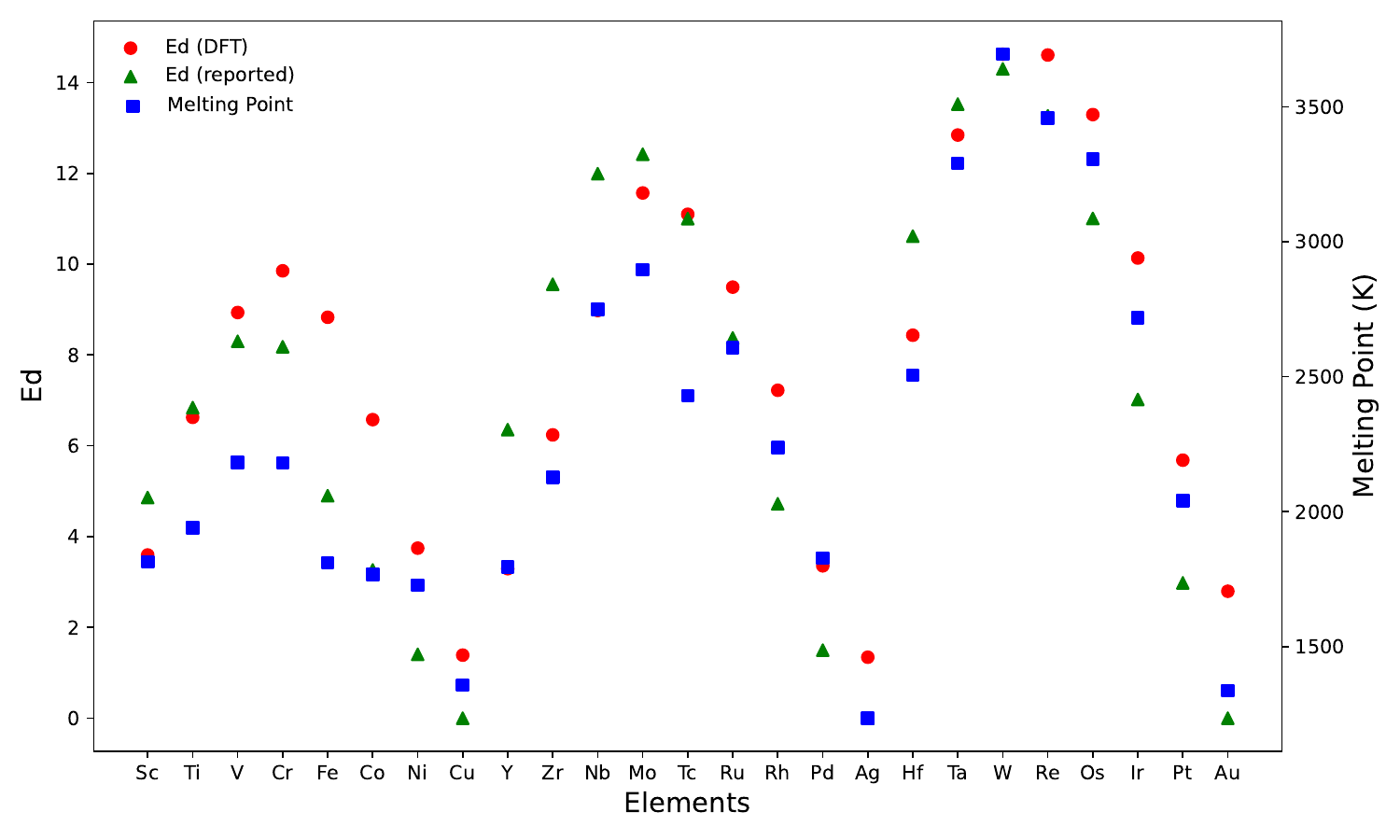}
    \caption{A comparison of d-electrons cohesive energy $E_{d}$ between reported values~\cite{harrison2012electronic} and DFT values for 3d,  4d and 5d elements. Higher cohesive energy leads to higher melting point (experimental values).}
\end{figure*}

\begin{figure*}
    \centering
    \includegraphics[width=0.9\textwidth]{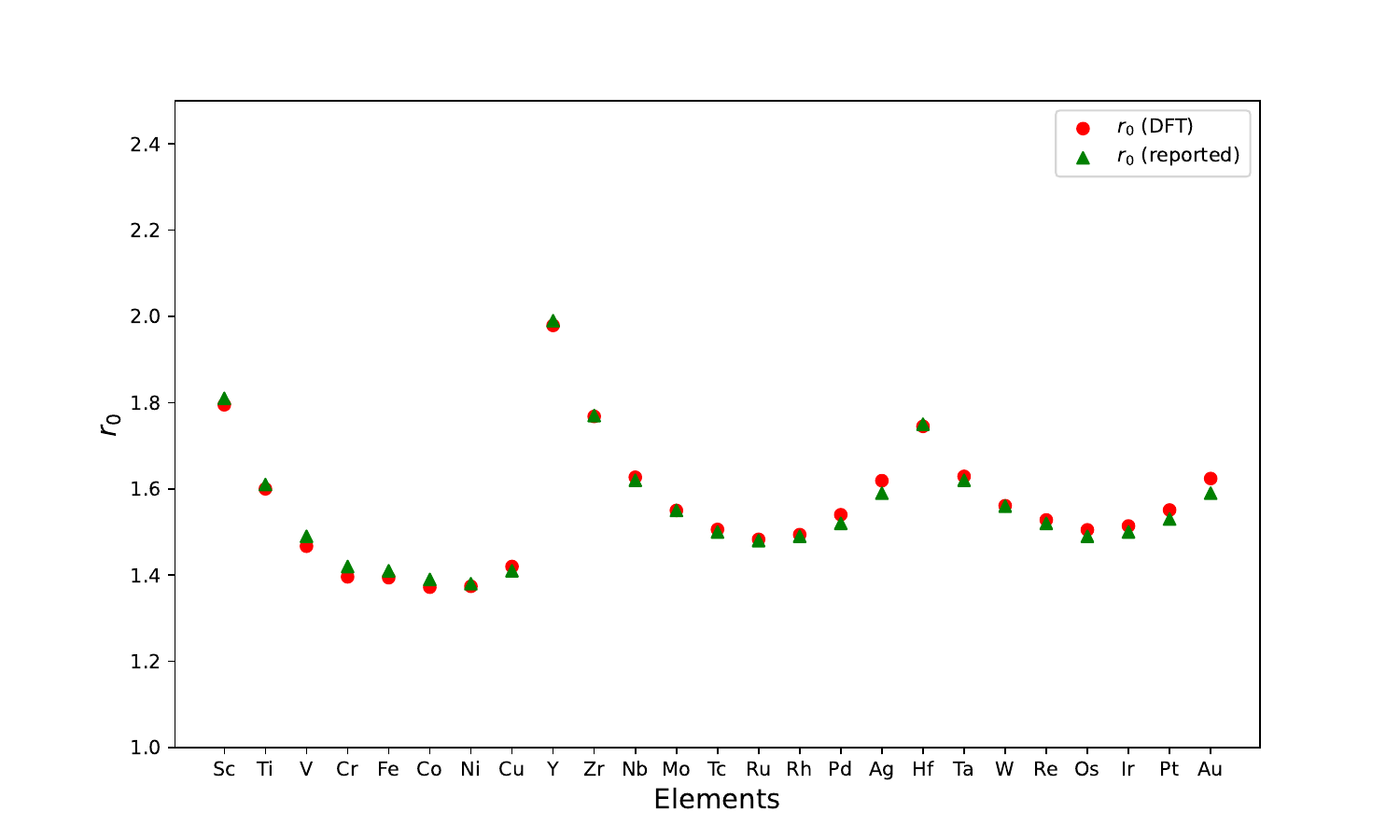}
    \caption{A comparison of Wigner seitz radius $r_{o}$ between reported values~\cite{harrison2012electronic} and DFT values for 3d,  4d and 5d elements.}
\end{figure*}

\begin{figure*}
    \centering
    \includegraphics[width=0.9\textwidth]{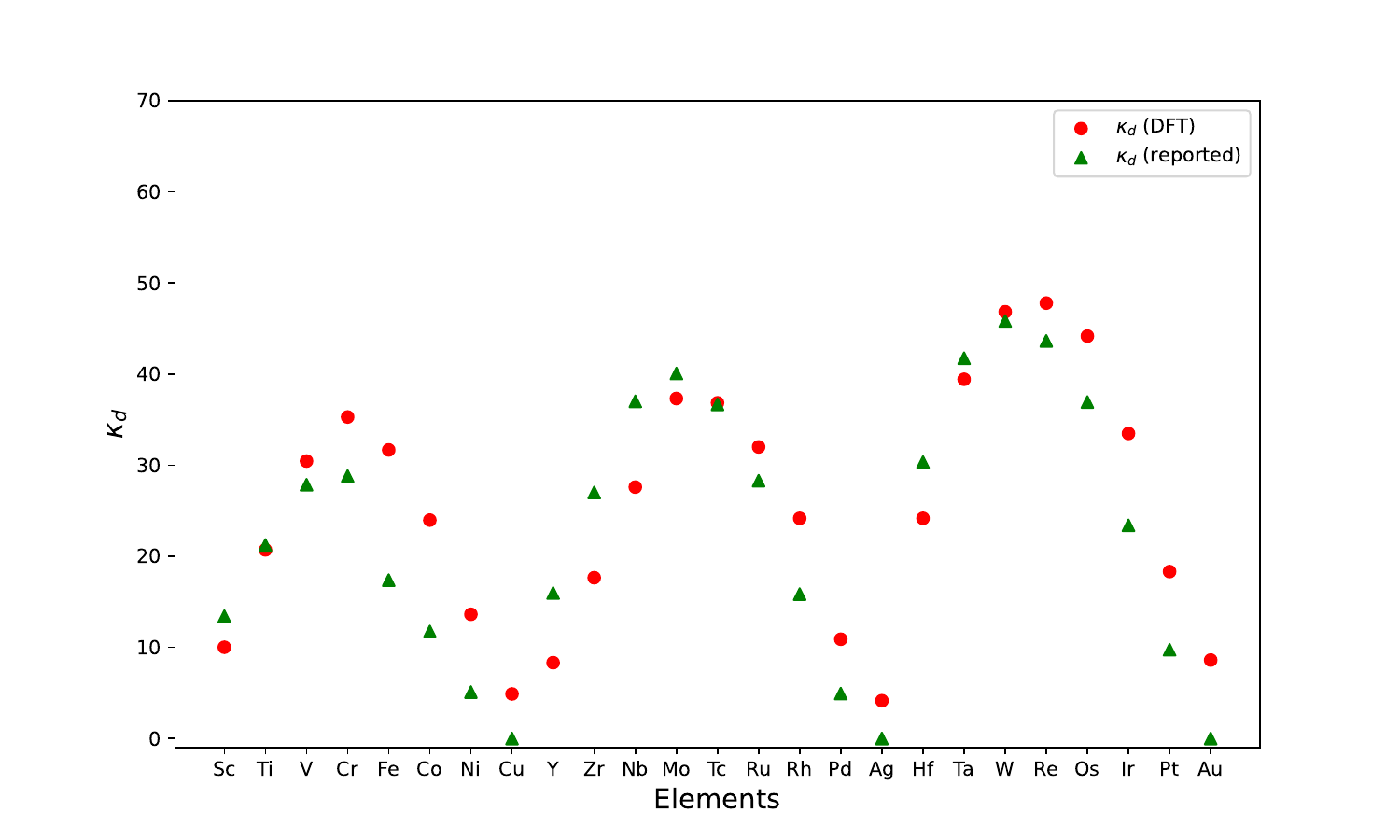}
    \caption{A comparison of d-electrons bulk modulus $\kappa_{d}$ between reported values~\cite{harrison2012electronic} and DFT values for 3d, 4d and 5d elements.}
\end{figure*}

\begin{figure*}
    \centering
    \includegraphics[width=1\textwidth]{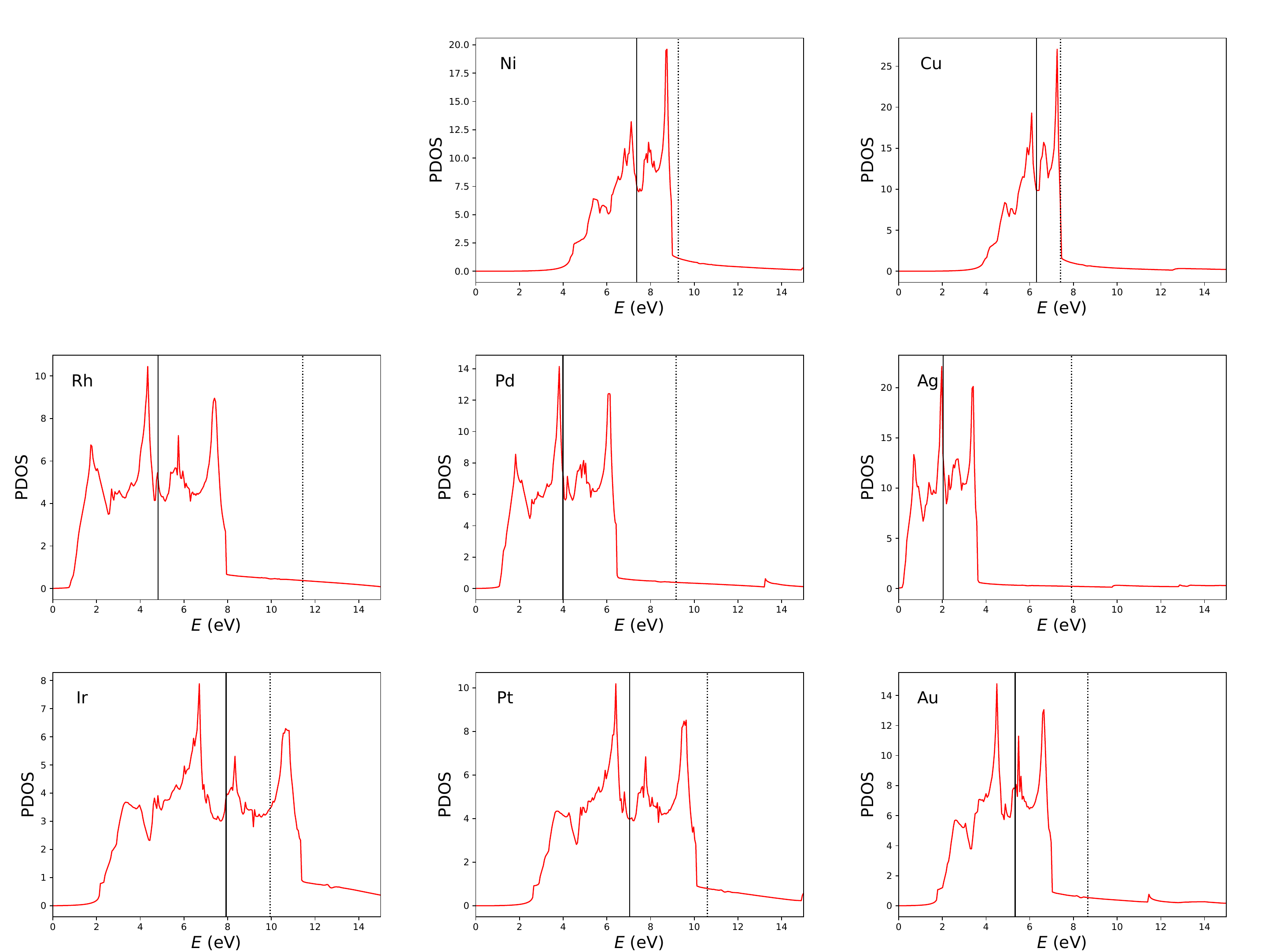}
    \caption{PDOS of d-orbitals for transition metals of 3d (upper row), 4d (middle row), and 5d (lower row). The solid black line indicates the center of the d-band ($\varepsilon_d$), and the dashed line represents the Fermi energy level ($E_F$).}
\end{figure*}

\begin{figure*}
    \centering
    \includegraphics[width=1\textwidth]{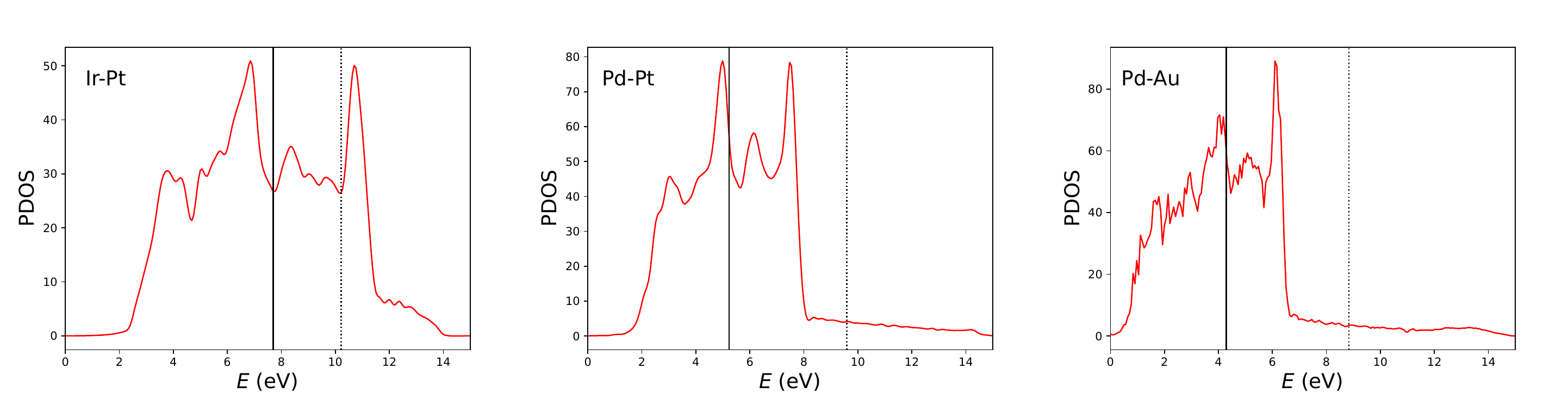}
    \caption{PDOS plots of d-orbitals for alloys of Ir$_{0.75}$Pt$_{0.25}$, Pd$_{0.75}$Pt$_{0.25}$, and Pd$_{0.75}$Au$_{0.25}$. The solid black line indicates the center of the d-band ($\varepsilon_d$), and the dashed line represents the Fermi energy level ($E_F$).}
    \label{}
\end{figure*}

\clearpage
\section{Prediction of Shear Modulus and SFE}

\subsection{Gaussian Process Regression}
Gaussian process regression (GPR) is a robust statistical model designed to learn the underlying function and its associated uncertainty. Gaussian process models estimate uncertainty that increases as one moves away from the training points. This feature quantifies uncertainty when using the surrogate model in future design tasks. To find the best model, we examine and compare three kernels: Radial Basis Function (RBF), Matern, and Exponential. We set RBF to set specific bounds with regulation parameters. An additional parameter is introduced by the Matern kernel, an extension of the RBF, to modify the function's smoothness. The Exponential kernel, which is comparable to the RBF but has a more straightforward shape, is also considered. Finding the best kernel for the model is the goal of the comparison, which considers factors like smoothness, adaptability, and performance in oversampled regions of the space. The Matern kernel produced the best model and the most consistent results out of the three kernels tested for GPRs, with test $R^2$ of 0.982, 0.992, and 0.992 for $G$, $\gamma_{ISF}$ and $\gamma_{ESF}$.

\subsection{Support vector regression}
Support vector regression (SVR) uses a subset of training data, that is, support vectors, in the final regression model by finding the optimal hyperplane. As a result, SVR models have improved memory efficiency and are less vulnerable to biases in the sampling of the training set. However, using these models makes quantifying uncertainty more difficult. Additionally, we test and compar four kernels, the polynomial, linear, sigmoid, and RBF, to choose the optimum SVR model with the help of grid search cross-validation. SVRs aim to maintain the residuals of all training data below a certain threshold, which lessens the temptation to sacrifice residuals on the minority of data, which results in RBF kernel being less prone to sample biasing and thus gives better results than GPR. The best SVR uses the RBF kernel and is trained with similar data points using Grid Search CV to estimate the best parameters, using which we obtain an $R^2$ score of 0.984, 0.994, and 0.996 for $G$, $\gamma_{ISF}$ and $\gamma_{ESF}$.
\begin{figure*}
    \centering
    \includegraphics[width=1\textwidth]{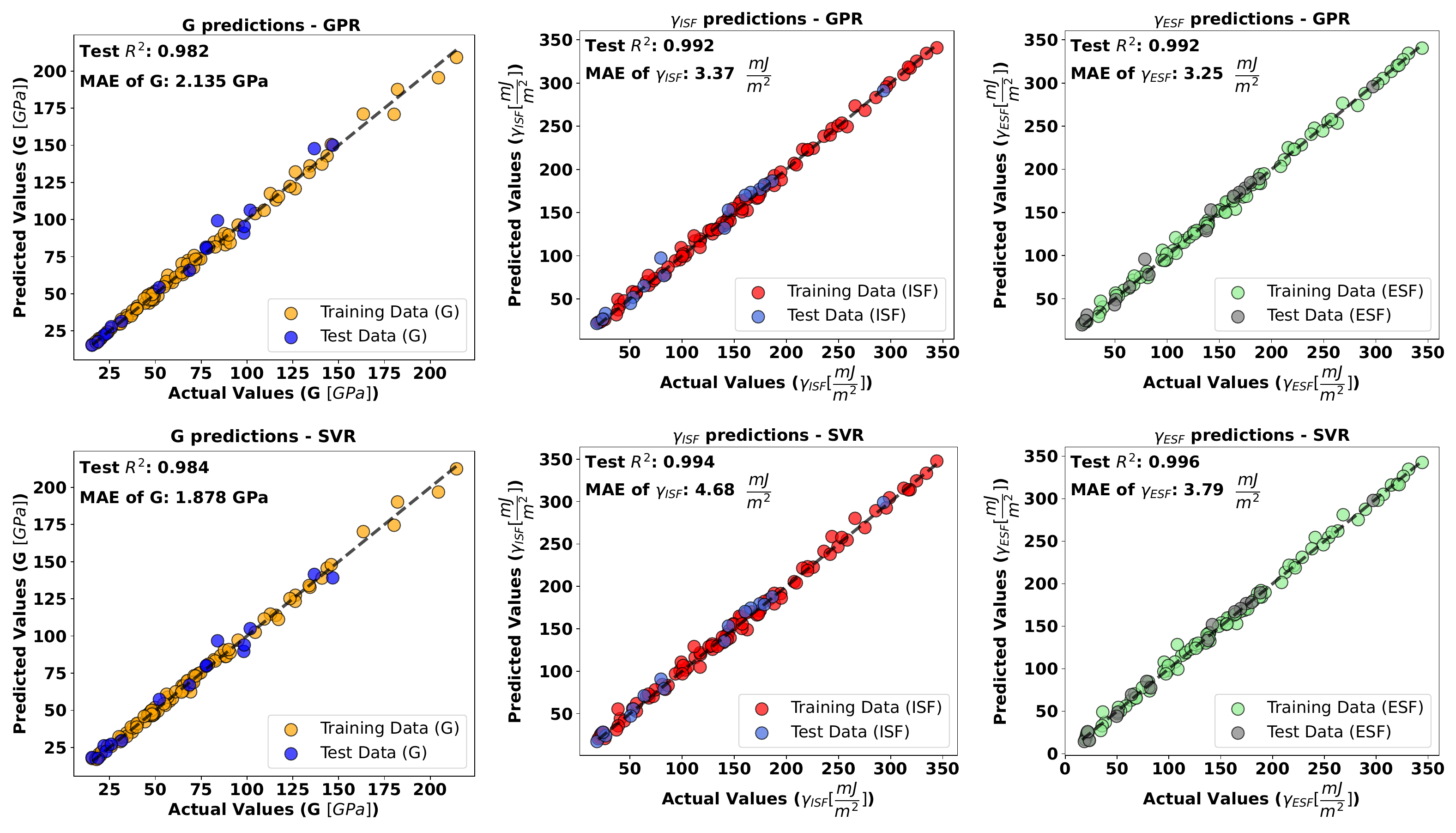}
    \caption{From left to right, the panels illustrate $G$, $\gamma_{ISF}$, and $\gamma_{ESF}$. Top row (Gaussian Process Regression) and bottom row (Support Vector Regression) represents the comparison of predicted vs actual values for training and test datasets.}
    \label{}
\end{figure*}

\begin{figure*}
    \centering
    \includegraphics[width=1\textwidth]{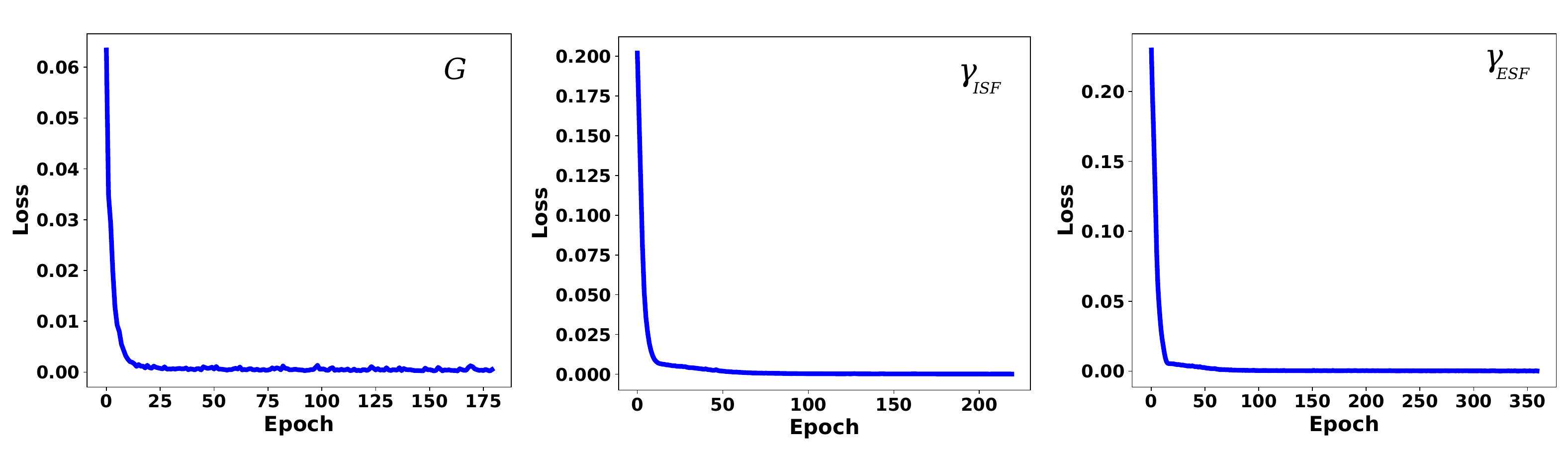}
    \caption{Epoch vs. loss plots depict the training process of DNNs for $G$, $\gamma_{ISF}$, and $\gamma_{ESF}$. The accuracy increases with the number of epochs, reaching a saturation point thereafter.}
    \label{}
\end{figure*}

\clearpage
\section{Feature importance analysis}
One thousand random trees with a maximum depth ranging from 1 to 100 terms are analyzed to understand the importance of features in a random forest. During the hyperparameter tuning process, we investigate each combination of hyperparameters using Grid-Search Cross-Validation, with different combinations of the number of estimators (random arrangement of parameters) and different maximum depths of branches (selected terms) of random forest trees. Then, we train the model using cross-validation on the different subsets of a combination of randomly selected parameters from the available data to minimize the mean square error of prediction. The feature importance is determined based on the contribution of each feature to the reduction in impurity (reduction in disorder or uncertainty in a data collection attained by decision tree splitting), i.e., minimizing the entropy along with minimizing the mean square error when branches of a random forest tree are split, as quantified by the formula:
\begin{equation}
    \text{Importance}(j) = \frac{\sum_{t}I(j \text{ is used to split in tree }t)\times \text{improvement in impurity}}{\text{Number of trees}}. 
\end{equation}
Here, $j$ represents the index of the feature, which takes the sum of over 1000 trees in the forest. The resulting importance scores provide insights into the relevance of each feature in making accurate predictions. The grid search identifies the best hyperparameters and the features contributing most to the model's accuracy. The selected terms include the first feature with the overall highest feature importance and other features from the decision tree that minimize the total number of terms and mean square error of predicting terms.
\bibliographystyle{ieeetr}
\bibliography{references}